\documentclass[article,notitlepage,twocolumn,preprintnumbers,superscriptaddress,nofootinbib]{revtex4-1}
\usepackage[utf8]{inputenc}
\usepackage[a4paper, total={7in, 9in}]{geometry}
\usepackage{amsmath}
\usepackage{amssymb}

\usepackage{enumerate}
\usepackage{amsmath}
\usepackage{tikz}
\usepackage[compat=1.1.0]{tikz-feynman}
\usepackage{feynmf}
\usepackage{slashed}
\usepackage{braket}
\usepackage{url}
\usepackage{natbib}
\usepackage{graphicx}
\usepackage{appendix}
\usepackage{tabularx}
\usepackage{multirow}

\usepackage{makecell}

\usepackage[pdftex,bookmarks,linktocpage,pdfpagelabels,plainpages=false,hyperfigures,linkcolor=blue,citecolor=blue]{hyperref} 
\hypersetup{colorlinks=true}

\usepackage{mathrsfs,amssymb}  
\usepackage{cancel}
\usepackage[normalem]{ulem}

\newcommand{\vb}[1]{\boldsymbol{#1}}
\newcommand{\polvec}{\vb\varepsilon}

\begin{document}
\bibliographystyle{apsrev4-1}
\title{Bragg-Primakoff Axion Photoconversion in Crystal Detectors}

\author{James B. Dent}
\affiliation{Department of Physics, Sam Houston State University, Huntsville, TX 77341, USA}

\author{Bhaskar Dutta}
\affiliation{Mitchell Institute for Fundamental Physics and Astronomy, Department of Physics and Astronomy, Texas A\&M University, College Station, TX 77845, USA}

\author{Adrian Thompson}
\affiliation{Mitchell Institute for Fundamental Physics and Astronomy, Department of Physics and Astronomy, Texas A\&M University, College Station, TX 77845, USA}

\begin{abstract}
Axions and axion-like pseudoscalar particles with dimension-5 couplings to photons exhibit coherent Primakoff scattering with ordered crystals at keV energy scales, making for a natural detection technique in searches for solar axions. We find that there are large suppressive corrections, potentially greater than a factor of $\mathcal{O}(10^3)$, to the coherent enhancement when taking into account absorption of the final state photon. This effect has already been accounted for in light-shining-through-wall experiments through the language of Darwin classical diffraction, but is missing from the literature in the context of solar axion searches that use a matrix element approach. We extend the treatment of the event rate with a heuristic description of absorption effects to bridge the gap between these two languages. Furthermore, we explore the Borrmann effect of anomalous absorption in lifting some of the event rate suppression by increasing the coherence length of the conversion. We study this phenomenon in Ge, NaI, and CsI crystal experiments and its impact on the the projected sensitivities of SuperCDMS, LEGEND, and SABRE to the solar axion parameter space. Lastly, we comment on the reach of multi-tonne scale crystal detectors and strategies to maximize the discovery potential of experimental efforts in this vein.
\end{abstract}

\preprint{MI-HET-804}

\maketitle

\section{Introduction}
Axions and axion-like particles (ALPs) - potentially long-lived pseudoscalars with weak couplings to the Standard Model (SM) that may have masses from the sub-eV to the GeV - are central features in the landscape of solutions to the strong CP problem~\cite{Peccei:1977hh,Wilczek:1977pj,Weinberg:1977ma,Preskill:1982cy,Abbott:1982af,Dine:1982ah,Battaglieri:2017aum}, dark matter problem~\cite{Marsh:2015xka,Arias:2012az, Duffy:2009ig,Adams:2022pbo}, and in the spontaneous breaking of generic global symmetries~\cite{Chikashige:1980ui,Svrcek:2006yi,Goodsell:2010ie}. In addition to being dark matter candidates, axion-like particles in the keV to sub-eV mass range produced in the sun are well motivated~\cite{Sikivie:1983ip,Raffelt:1996wa}. Searches were carried out by several experimental collaborations by looking for $a \to \gamma$ Primakoff conversion in solid crystal detectors, including DAMA~\cite{Bernabei:2004fi} (NaI), CUORE~\cite{Li:2015tyq, Li:2015tsa} (TeO$_2$), Edelweiss-II~\cite{Armengaud_2013}, SOLAX~\cite{PhysRevLett.81.5068}, COSME~\cite{COSME:2001jci}, CDMS~\cite{CDMS:2009fba}, and Majorana~\cite{Majorana:2022bse} (Ge). Other upcoming experiments like SuperCDMS~\cite{SuperCDMS:2022kse}, LEGEND~\cite{LEGEND:2021bnm}, and SABRE~\cite{SABRE:2018lfp} are projected to greatly expand coverage over the axion parameter space and test QCD axion solutions to the strong CP problem in the eV mass range. These experiments aim to take advantage of coherence in the conversion rate when axions satisfy the Bragg condition, enhancing the detection sensitivity by orders of magnitude relative to incoherent scattering.

Searching for solar axions via their coherent conversion in perfect crystals was first treated by Buchm\"uller \& Hoogeveen~\cite{Buchmuller:1989rb} using the Darwin theory of classical X-ray diffraction under the Bragg condition~\cite{warren}. The authors also alluded to potential enhancements in the signal yield when one considers the symmetrical Laue-case of diffraction for the incoming ALP waves. Yamaji \textit{et al.}~\cite{Yamaji:2017pep} treated this case thoroughly for the 220 plane of cubic crystals, also using the classical theory, and included the effect of anomalous absorption, also known as the Borrmann effect. It was shown by these authors that an enhancement to the signal yield was possible, replacing the Bragg penetration depth ($L_\textrm{bragg} \sim 1$~$\mu$m) with the Borrmann-enhanced attenuation length (ranging from 10~$\mu$m all the way to centimeter scales).

The effect of anomalous absorption of X-rays was first shown by Borrmann~\cite{borrmann1954}, and theoretically explained by Zachariasen~\cite{zachariasenBook,zachariasenPNAS} and other later authors (Battermann~\cite{batterman1,batterman2}, Hirsch~\cite{Hirsch:a00588}). A quantum mechanical treatment was offered by Biagini~\cite{PhysRevA.42.3695,PhysRevA.44.645} in which the Borrmann effect was explained by the interference of statistical ensembles of the so-called $\ket{\alpha}$ and $\ket{\beta}$ Bloch waves. There have been numerous modern studies that utilize the Borrmann effect, notably as in photon-photon dissipation on Bragg-spaced arrays of superconduncting qubits~\cite{Poshakinskiy_2021}, and in measuring quadrupole transitions in X-ray absorption spectra~\cite{pettifer_collins_laundy_2008}.

Now, the calculation of the event rates expected for the Primakoff conversion of solar axions coherently with a perfect crystal was treated in a more traditional, particle physics-based approach in refs.~\cite{Cebrian:1998mu,Bernabei:2001ny,Li:2015tsa} and it was applied to derive many of the constraints set by crystal-based solar axion experiments including DAMA, CUORE, Edelweiss-II, SOLAX, COSME, CDMS, and Majorana Demonstrator~\cite{Bernabei:2004fi,Li:2015tyq,Armengaud_2013,PhysRevLett.81.5068,COSME:2001jci,CDMS:2009fba,Majorana:2022bse}. However, absorption effects in Bragg and Laue case diffraction were not considered in refs.~\cite{Cebrian:1998mu,Bernabei:2001ny,Li:2015tsa}; indeed, when comparing the event rates between these references and those presented in light-shining-through-wall (LSW) experiments, which used the classical Darwin theory approach (e.g. ref.~\cite{Buchmuller:1989rb} and more recently ref.~\cite{Yamaji:2017pep}), there is a clear inconsistency. While the event rates in the LSW literature only consider the coherent volume of the crystal up to the relevant attenuation length ($\lambda \sim 1$ $\mu$m in the Bragg diffraction case or $\lambda \lesssim 100$ $\mu$m in the Laue-case), the solar axion searches have considered the whole volume of the crystal to exhibit coherence. In this work, we show that such effects reduce the expected event rates potentially up to the $\mathcal{O}(10^3)$ level depending on the assumed crystal size (and therefore, the assumed coherent volume enhancement) and material. Although this may impact the existing sensitivities set by solar axion searches in solid crystals, measures can be taken to optimize suppression of the event rate due to absorption effects and recover some or potentially all of the coherent volume.

In \S~\ref{sec:derivation} we re-derive the event rate formula for solar axion Primakoff scattering under the Bragg condition, and in \S~\ref{sec:borrmann} we discuss the anomalous enhancement to the absorption length under the Borrmann effect and numerically estimate the level of suppression in the coherent sum. In \S~\ref{sec:event_rates} we write down the event rates for a perfect crystal exposed to the solar axion flux with and without the absorption effects and discuss the relevant phenomenology. In \S~\ref{sec:sensitivities} we project the impact on sensitivities with and without absorption effects for SuperCDMS, LEGEND-200, LEGEND-1000, SABRE, and multi-tonne benchmark detector setups and discuss possibilities to restore sensitivity from coherence in \S~\ref{sec:restoring_coh}. Finally, in \S~\ref{sec:conclude} we conclude and discuss further work.


\section{Coherence and Absorption}
\label{sec:derivation}
In order to show how photon absorption in coherent Bragg-Primakoff scattering affects the event rate, it is worth going through a pedagogical review of what we mean by coherent scattering and first assume that no absorption takes place. For the reader who is familiar with coherence in neutrino scattering, please refer to the approach illustrated by Bednyakov and Naumov~\cite{Bednyakov:2021lul} in which coherent neutrino-nucleus scattering is calculated by taking a sum over $N$ scattering centers in a nucleus. 

Let $f(\vec{k},\vec{k}^\prime)$ be the Primakoff scattering matrix element for a single atomic target, for an incoming ALP 3-momentum $\vec{k}$ and outgoing $\gamma$ 3-momentum $\vec{k}^\prime$. Written in terms of the atomic form factor $F_A$,
\begin{equation}
    f = \mathcal{M}_\textrm{free} F_A (q)
\end{equation}
where $\mathcal{M}_\textrm{free}$ is the single-atomic scattering amplitude, $q$ is the momentum transfer, with the angle of scattering defined by $\vec{k} \cdot \vec{k}^\prime = E_\gamma k \cos2\theta$, averaged over spins and taken in the limit $k \gg m_a$, $m_N \gg k,E_\gamma$~\cite{Tsai:1986tx},
\begin{equation}
    |\braket{\mathcal{M}_\textrm{free}}|^2 = \dfrac{8 e^2 g_{a\gamma}^2}{q^4} E_\gamma^2 m_N^2 k^2 \sin^2 2\theta 
\end{equation}
for a nuclear mass $m_N$. The real atomic scattering form factor can be taken from ref.~\cite{RevModPhys.46.815} which is defined such that $F_A(0) = Z$;
\begin{equation}
    F_A(q) = \dfrac{Z r_0^2 q^2}{1 + r_0^2 q^2}
\end{equation}
for atomic number $Z$ and screening constant parameterization $r_0=184.15 e^{-1/2} Z^{-1/3} / m_e$, where $m_e$ is the electron mass. 

Similarly, we sum over the $N$ scattering centers in a crystal;
\begin{equation}
    \mathcal{M}(\vec{k},\vec{k}^\prime) = \sum_{j=1}^N f_j(\vec{k},\vec{k}^\prime) e^{i(\vec{k}^\prime - \vec{k})\cdot \vec{r}_j}
\end{equation}
where $e^{i(\vec{k}^\prime - \vec{k})\cdot \vec{r}_j}$ is a phase factor that comes from assuming plane wave solutions for the in and out states. This assumption is key; for atomic scattering in vacuum, the eigenstates of the final state photon should be a spectrum of plane waves. 

If we square the total matrix element, we get
\begin{equation}
     \mid\mathcal{M}(\vec{k},\vec{k}^\prime)\mid^2 = \sum_{i=1}^N \mid f_i\mid^2 + \sum_{j\neq i}^N \sum_{i=1}^N f_j^\dagger f_i e^{-i\vec{q}\cdot(\vec{r}_i - \vec{r}_j)}
\end{equation}
taking $\vec{q} \equiv \vec{k} - \vec{k}^\prime$. The first (diagonal) term is the \textit{incoherent} piece, while the second term is usually suppressed by the average destructive interference of the phase factors. Using the Laue diffraction condition~\cite{warren}, $\vec{q}\cdot(\vec{r}_i - \vec{r}_j) = 2\pi n$ for $n\in \mathbb{Z}$, then the phase factor in the exponential goes to one and the scattering is \textit{coherent}. In this limit, the diagonal term is subdominant and the final matrix element squared tends to $\mathcal{M}^2 \to N^2 f^2$ and we have full coherence. See appendix~\ref{app:derivation} for a derivation of the event rate in full with this approach.

Now consider interactions of the final state $\gamma$ with the crystal lattice, including the absorption and scattering effects. Pragmatically, we modify the plane wave solutions of the final state photon to that of one in a dielectric medium,
\begin{equation}
\label{eq:dielectric}
    \vec{k}^\prime \to \Bar{n} \vec{k}^\prime, \, \, \, \Bar{n} = n - i \kappa,
\end{equation}
where $\bar{n}$ is the complex index of refraction with real part $n$ and imaginary part $\kappa$. Making this modification, we have
\begin{align}
\label{eq:abs_phase}
    e^{i \bar{n} \vec{k}^\prime \cdot (\vec{r}_i - \vec{r}_j)} &\to e^{i n \vec{k}^\prime \cdot (\vec{r}_i - \vec{r}_j)} e^{-\frac{\mu}{2} |\hat{k}^\prime \cdot (\vec{r}_j - \vec{r}_i)|},
\end{align}
The absorption coefficient $\mu$ (which can also be expressed in terms of attenuation length or mean free path $\lambda = 1\mu$) is related to the imaginary part of the index of refraction through $\mu \equiv 2 \kappa |\vec{k}|$. Conceptually, this factor encodes the effect of a reduced coherent interference amplitude between any two scattering centers, since a photon plane wave sourced at one scattering center will have been attenuated after reaching another scattering center.

We note that Eq.~\ref{eq:dielectric} and Eq.~\ref{eq:abs_phase} are heuristic modifications, since the attenuated plane wave solution is not a true eigenstate of the interaction Hamiltonian, but rather a simple ansatz made to estimate the phenomenology of absorption. For further convenience, we use $z_{ij} \equiv \mid\hat{k^\prime}\cdot(\vec{r}_i-\vec{r}_j)\mid$ and $\lambda = 1/\mu$. We then have
\begin{align}
     \mid\mathcal{M}(\vec{k},\vec{k}^\prime)\mid^2 &= \sum_{i=1}^N \mid f_i\mid^2 \nonumber \\
     +& \sum_{j\neq i}^N \sum_{i=1}^N f_j^\dagger f_j e^{-i\vec{q}\cdot(\vec{r}_i - \vec{r}_j)}e^{-z_{ij}/(2\lambda)}
\end{align}
After using the Laue diffraction condition $\vec{q}\cdot(\vec{r}_i - \vec{r}_j) = 2\pi n$ and several manipulations of the sum, we find that
\begin{align}
\label{eq:abs_volume}
    \mid\mathcal{M}(\vec{k},\vec{k}^\prime)\mid^2 &\gtrsim f^\dagger f\sum_{j\neq i}^N \dfrac{\lambda L_x L_y N}{V} \nonumber \\
    &\gtrsim f^\dagger f N^2 \dfrac{\lambda}{L_z}
\end{align}
Comparing the proportionailty in Eq.~\ref{eq:abs_volume} to the usual result $\propto N^2$, we see that the coherent volume is $V \times \lambda / L_z$, and the total scattering rate is suppressed by a factor $\lambda/L_z$, and is now more consistent with Darwin theory calculations~\cite{Buchmuller:1989rb, Yamaji:2017pep}.

This inequality above is strictly a lower limit because, as we will show in \S~\ref{sec:borrmann}, the suppression to the coherent sum by the absorptive sum, which we label as $I$,
\begin{equation}
\label{eq:abs_sum}
   I \equiv \sum_{j\neq i}^N \sum_{i=1}^N e^{-z_{ij}/(2\lambda)},
\end{equation}
may be mitigated under certain conditions. Therefore, the suppression factor $\lambda / L_z$ serves as a pessimistic guiding estimate, but in principle we should compute the sum in Eq.~\ref{eq:abs_sum} explicitly.
\begin{figure}[h!]
    \centering
    \includegraphics[width=0.23\textwidth]{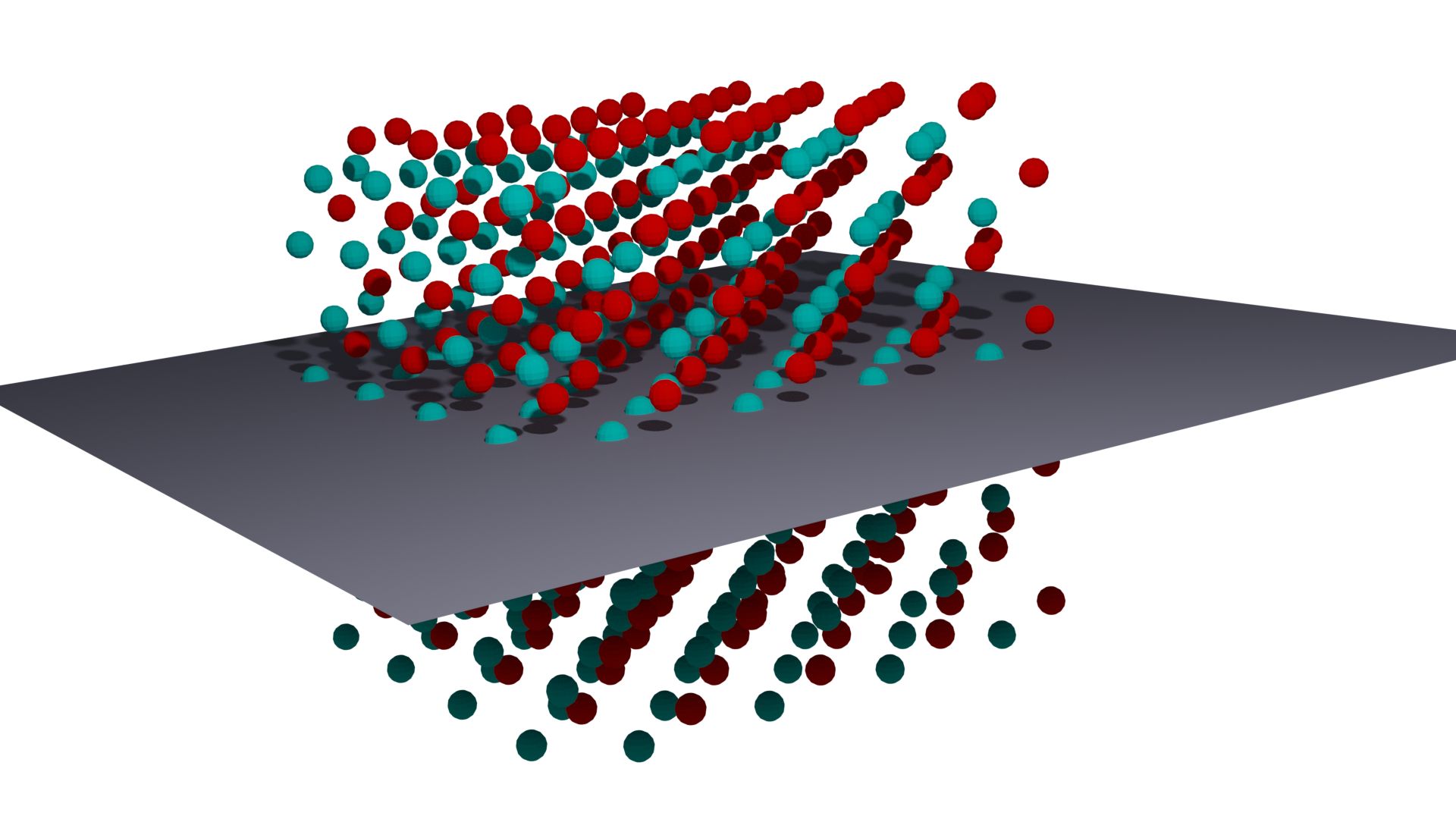}
    \includegraphics[width=0.23\textwidth]{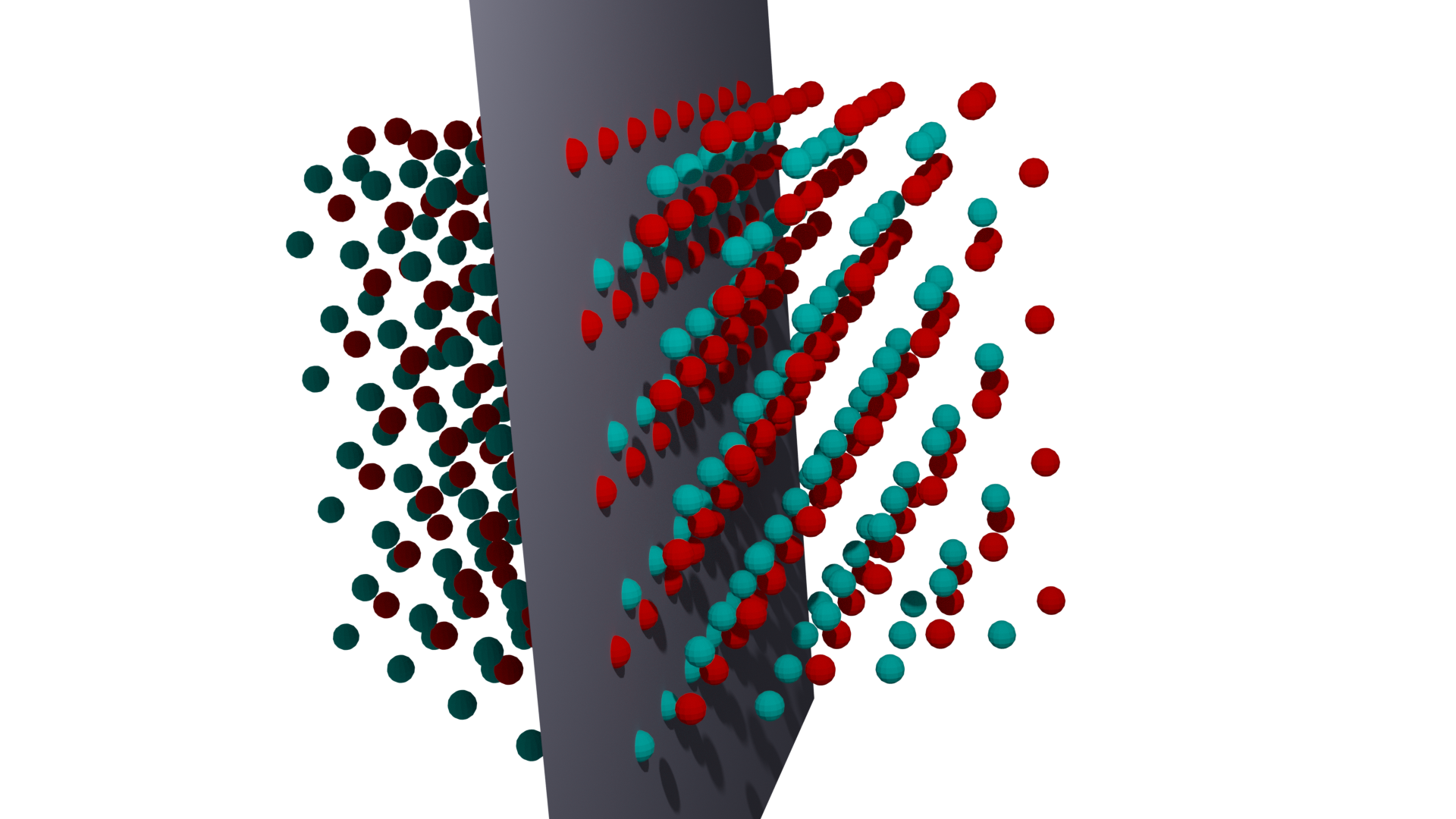}\\
    \includegraphics[width=0.23\textwidth]{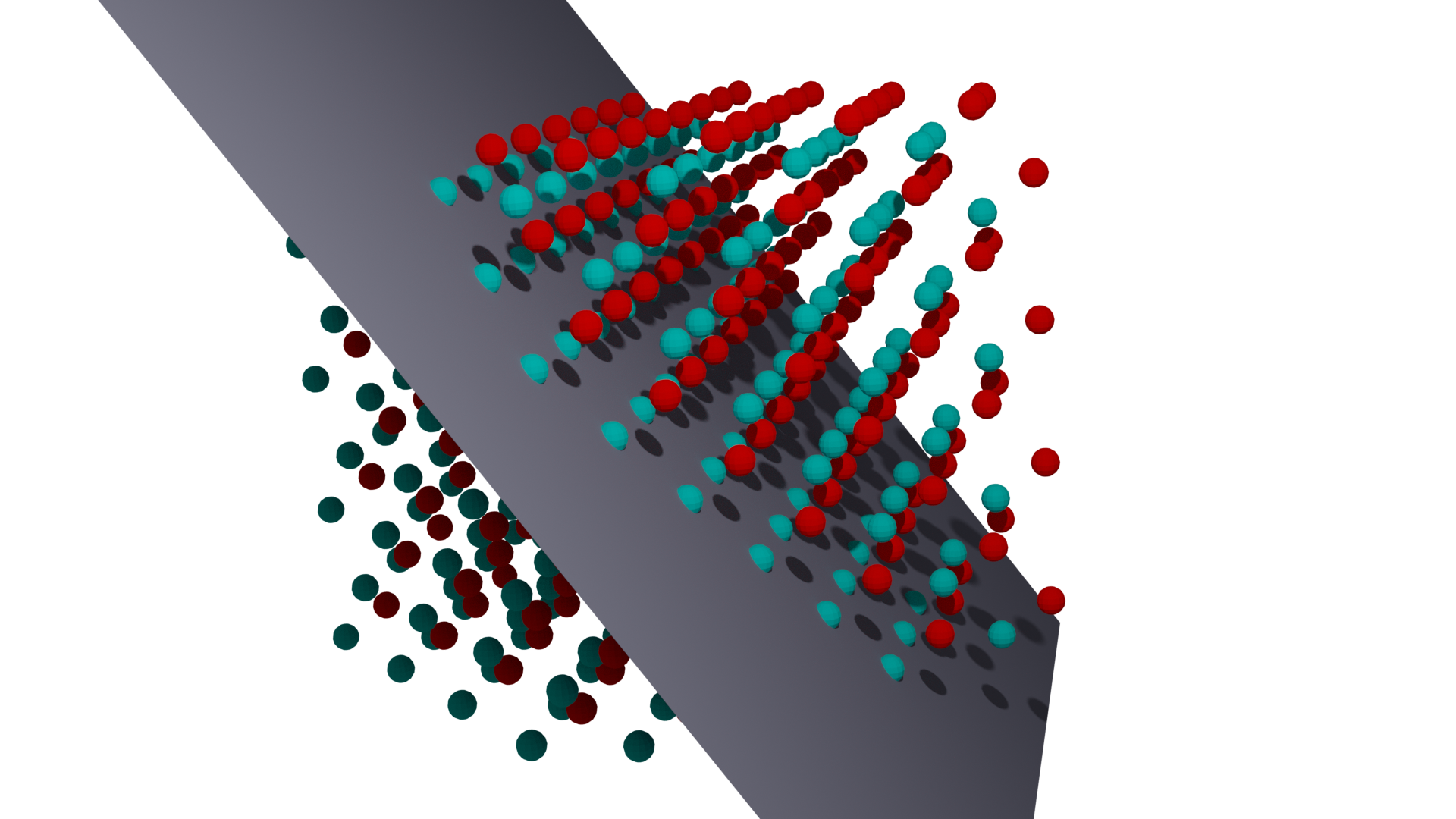}
    \includegraphics[width=0.23\textwidth]{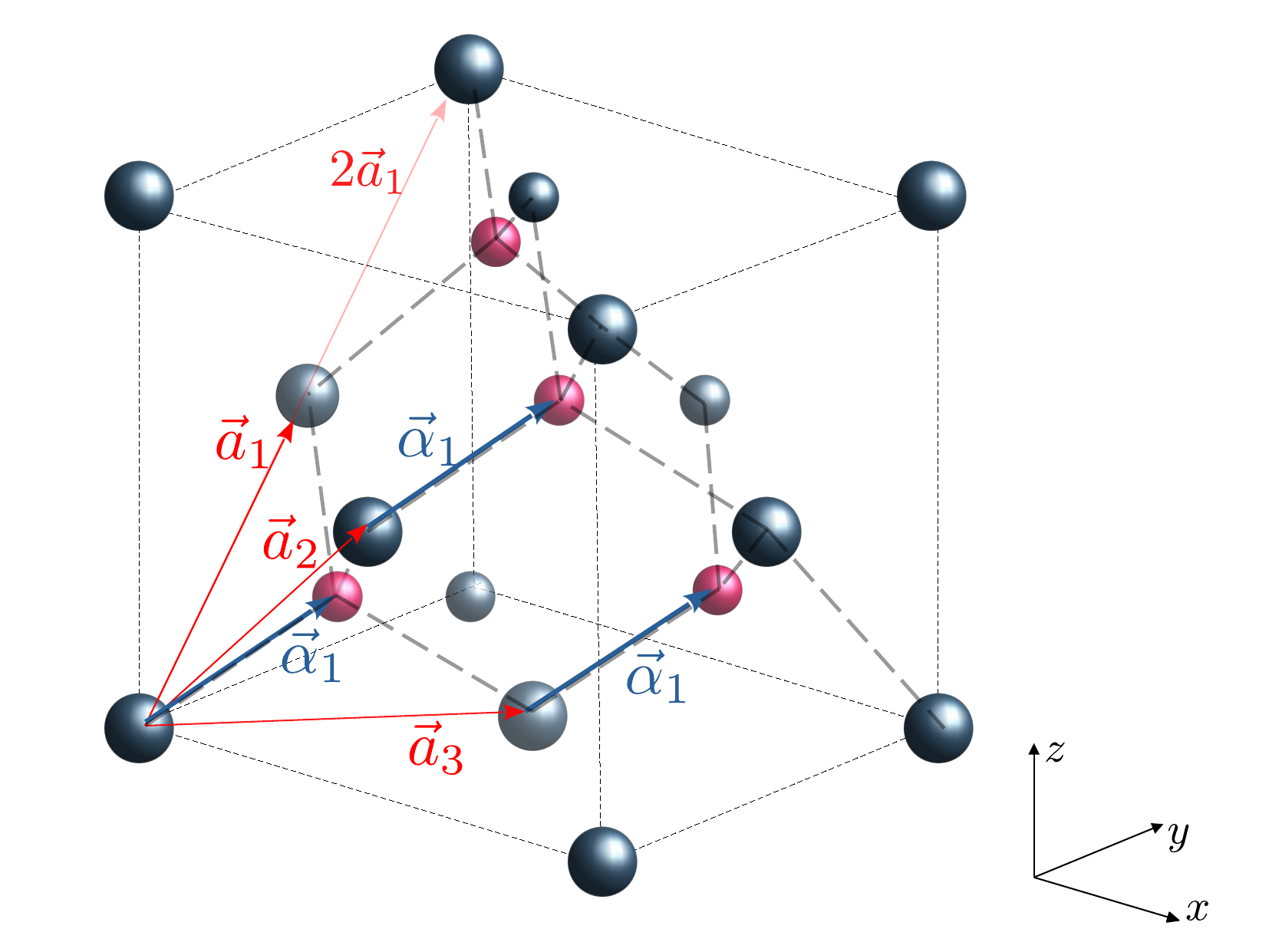}
    \caption{Crystallographic planes cut by the $hkl=220$, 224, and 333 reciprocal lattice vectors in the FCC lattice, and their basis vectors.}
    \label{fig:csi_planes}
\end{figure}

\section{Anomalous Absorption and the Borrmann Effect}
\label{sec:borrmann}
The suppression to the event rate can be alleviated by considering the anomalous enhancement to the absorption depth or mean free path $\lambda$, which, in crystallographic diffraction, is not strictly proportional to the the inverse photon cross section multiplying into the material number desnsity, $1/(n\sigma)$.

Take for instance ref.~\cite{Yamaji:2017pep} in which the authors have found that for the Laue-case conversion of ALPs, the attenuation length is modified as
\begin{equation}
    L_\text{att} \to L_{\alpha / \beta} \equiv 2L_{\text{att},\alpha / \beta}\bigg(1 - \exp\bigg(-\dfrac{L}{2L_{\text{att},\alpha / \beta}} \bigg) \bigg) 
\end{equation}
where $L_{\text{att},\alpha / \beta} = \dfrac{L_\text{att}}{1 \mp \epsilon}$ and $\epsilon$ is a ratio involving the imaginary parts of the scattering form factor. These modifications come from the anomalous dispersion or anomalous absorption effect, or the Borrmann effect. It is an effect that occurs for so-called ``Bloch waves" $\alpha$ and $\beta$ that form in the crystal, discussed further in refs.~\cite{PhysRevA.42.3695, PhysRevA.44.645}.

The total scattering form factor can be decomposed into the real and imaginary parts~\cite{chantler_NIST};
\begin{equation}
    f = f^0 + \Delta f^\prime + i \Delta f^{\prime\prime}
\end{equation}
where $f^0$ is the atomic form factor, usually given as the Fourier transform of the charge density;
\begin{equation}
    f^0(q) \equiv \int d^3 \vec{x} \rho(\vec{x}) e^{i \vec{q}\cdot\vec{x}}
\end{equation}
The second term in the real part of the form factor is the anomalous form factor $\Delta f^\prime$, and $\Delta f^{\prime\prime}$ is the imaginary part of the form factor associated with absorption. From Batterman~\cite{batterman1, batterman2}, the anomalous absorption due to the Borrmann effect modifies the absorption coefficient $\mu_0 = 1/\lambda$ as
\begin{equation}
\label{eq:anomalous_depth}
   1/\lambda = \mu_\text{eff} = \mu_0 \bigg[1 - \dfrac{F^{\prime\prime}(hkl)}{F^{\prime\prime}(000)}\bigg]
\end{equation}
Here $F^{\prime\prime}(hkl)$ is the combination of structure function and imaginary form factor, $F^{\prime\prime}(hkl) = S(hkl) \Delta f^{\prime\prime}$. The ratio in the second term of the expression is the Borrmann parameter, usually denoted as $\epsilon$\footnote{In ref.~\cite{Yamaji:2017pep}, they use $\kappa$.}. More explicitly, studies by Wagenfield have related the Borrmann parameter to the quadrupole photoelectric cross section~\cite{wagenfield1986,PhysRev.144.216,persson_efimov};
\begin{equation}
    \epsilon \equiv D \bigg(1 - 2 \sin^2\theta_B \frac{\sigma^Q}{\sigma_\textrm{PE}} \bigg) \frac{|S(h,k,l)|}{|S(0,0,0)|}
\end{equation}
where $D$ is the Debye-Waller factor accounting for thermal vibrations in anomalous absorption, $D = e^{-B s^2}$ where $s = \sin\theta / \lambda$ and $B$ is a temperature-dependent constant. The Debye-Waller factors for cryogenic temperatures can be found in ref.~\cite{Peng96} as well as fits to $\Delta f^{\prime\prime}$ for several pure materials of interest. Equivalently, we can express the Borrmann factor in terms of the imaginary form factor $\Delta f^{\prime\prime}$ and the quadrupole form factor $\Delta f^{\prime\prime}_Q$ (which obeys the selection rules $\ell = \ell^\prime \pm 2$);
\begin{equation}
    \epsilon \equiv D \bigg(1 - 2 \sin^2\theta_B \frac{\Delta f^{\prime\prime}_Q}{\Delta f^{\prime\prime}} \bigg) \frac{|S(h,k,l)|}{|S(0,0,0)|}
\end{equation}
and $\Delta f^{\prime\prime}$ is more explicitly written as~\cite{wagenfield1986}
\begin{equation}
    \Delta f^{\prime\prime} = \sum_{\ell^\prime,m^\prime}  \sum_{n,\ell,m} \dfrac{\pi \hbar^2}{m_e} \bigg| \int \psi_f^*(\vb{r}) \polvec_0 \cdot \vb{\nabla} e^{i \vb{k}\cdot\vb{r}} \psi_i(\vb{r}) d^3 r \bigg|^2
\end{equation}
While fits to this form factor can be found in ref.~\cite{Peng96}, we can also usefully relate it to the vectorial form factor defined in ref.~\cite{Catena:2019gfa} and calculated using the \texttt{DarkARC} (Python) or \texttt{DarkART} (C++) codes;
\begin{equation}
    \Delta f^{\prime\prime}(k) =\pi \hbar^2 m_e |\vb{f}_{1\to2}(k)|_\texttt{DarkARC}^2
\end{equation}
For more discussion and example functional forms of the Borrmann parameter, see appendix~\ref{app:darkarc}.

While a dedicated study of the Borrmann parameter would require the calculation of the photoelectric quadrupole cross section $\sigma^Q$, Borrmann parameters for germanium crystal are already reported in the literature. We use the form factors derived in ref.~\cite{batterman1} to estimate the Borrmann effect for each reciprocal lattice plane, giving us an anomalous attenuation length along the direction of travel of photons inside the detector $I(\vec{k},\vec{G})$. We tabulate these and the corresponding values of $\epsilon$ in Table~\ref{tab:borrmann} and plot the Borrmann parameters for Ge, Si, CsI, and NaI crystals in Fig.~\ref{fig:borrmann_by_mat}.
\begin{table}[h]
\centering
  \begin{tabularx}{\columnwidth}{|X|X|X|X|X|}
  \hline
    \multirow{2}{*}{$(hkl)$} &
      \multicolumn{2}{c}{Ge} &
      \multicolumn{2}{c|}{Si} \\
    & $\epsilon$ & $\lambda$ [$\mu$m]& $\epsilon$& $\lambda$ [$\mu$m] \\
    \hline
111 & 0.69 & 34.2 &0.68 & 96.26 \\
220 & 0.95 & 229.69 & 0.91 & 357.44 \\
131 & 0.66&31.08&0.63&82.62 \\
400 & 0.91 & 122.95 & 0.85 & 209.55 \\
133 & 0.64 & 28.77 & 0.59 & 74.85 \\
422 & 0.88 & 86.13 & 0.81 & 158.64 \\
440 & 0.85 & 67.99 & 0.77 & 131.92 \\
\hline
  \hline
    \multirow{2}{*}{$(hkl)$} &
      \multicolumn{2}{c}{NaI} &
      \multicolumn{2}{c|}{CsI} \\
    & $\epsilon$ & $\lambda$ [$\mu$m]& $\epsilon$& $\lambda$ [$\mu$m] \\
    \hline
200 & 0.92 & 181 & 0.53 & 27.03 \\
220 & 0.86 & 104.3 & 0.27 & 17.47 \\
222 & 0.82 & 76.84 & 0.13 & 14.55 \\
400 & 0.77 & 62 & 0.04 & 13.28 \\
420 & 0.73 & 52.48 & 0 & 12.66 \\
440 & 0.62 & 37.1 & -0.05 & 12.13 \\
600 & 0.58 & 34.12 & -0.05 & 12.11 \\
\hline
  \end{tabularx}
    \caption{Anomalous coefficients $\epsilon(hkl)$ and attenuation lengths $\lambda(E_\gamma, hkl)$ in Ge, Si (diamond cubic), NaI, and CsI (FCC) due to the Borrmann effect were computed using the imaginary form factor with $E_\gamma = 3$ keV. At this energy the normal attenuation length is $\sim 10$ $\mu$m. See also Fig.~\ref{fig:borrmann_by_mat}.}
    \label{tab:borrmann}
\end{table}

\begin{figure}
    \centering
    \includegraphics[width=0.48\textwidth]{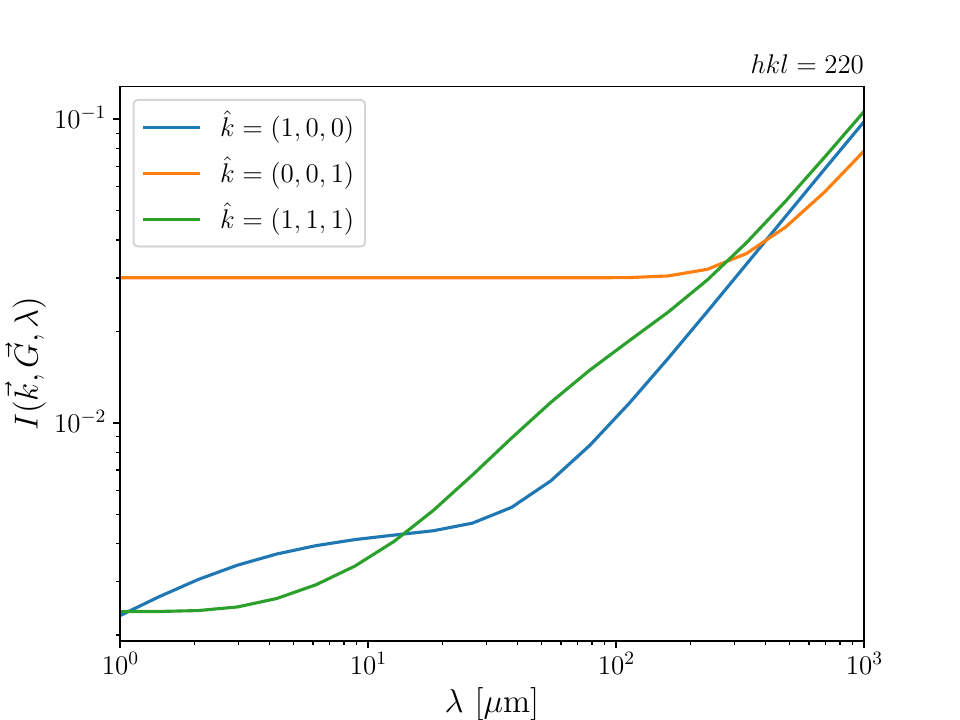}
    \caption{The absorption factor $I(\vec{k}, \vec{G})$ as a function of the mean free path $\lambda = 1/\mu$ for a crystal of cubic volume with side length 5 cm.}
    \label{fig:abs_factor}
\end{figure}

The absorptive part of the coherent sum that remains after the Laue condition is met is
\begin{equation}
   I(\vec{k},\vec{G}) \equiv \sum_{j\neq i}^N \sum_{i=1}^N e^{-\frac{(\vec{k}-\vec{G})}{|\vec{k}-\vec{G}|}\cdot(\vec{r}_i - \vec{r}_j)/(2\lambda)}
\end{equation}
which, when the Bragg condition is met, is strictly a function of $\vec{k}$ and $\vec{G}$ since the mean free path $\lambda$ can be related via Eq.~\ref{eq:anomalous_depth}. Taking the Ge lattice as an example, with lattice constant $d = 5.657$  \AA, we evaluate $I(\vec{k},\vec{G})$ numerically by constructing a lattice of $N$ Ge atoms. Since computing the full sum for a real crystal of centimeter length scale would require a huge number of evaluations $(\propto N^2)$, we take a sparse sampling of $N$ atoms across the physical crystal volume such that the sum is computationally feasible. The sum can then be evaluated in increments of increasing $N$ to test for convergence. We find that a lattice of around $N\simeq 10^4$ atoms in a cubic geometry is enough to obtain a convergent error of around 5\%. Some evaluations of $I(\vec{k},\vec{G})$ as a function of varying mean free path $\lambda$ are shown in Fig.~\ref{fig:abs_factor} for several choices of scattering planes $\vec{G}$ and incoming wavevectors $\vec{k}$.

One interesting phenomenon that can be seen in Fig.~\ref{fig:abs_factor} is that there are certain choices of $\vec{k}^\prime = \vec{k} - \vec{G}$ such that $\vec{k}^\prime \cdot (\vec{r}_i - \vec{r}_j) = 0$. In this special circumstance, while many of the terms in the coherent sum will tend to zero with decreasing $\lambda$, the terms where this dot product is zero will survive. What this means physically is that the plane in which $\vec{r}_i - \vec{r}_j$ lies will avoid the decoherence from absorption as long as it remains orthogonal to $\vec{k}^\prime$. This relation can be made more apparent by considering the dot product under the Bragg condition;
\begin{align}
    \hat{k}^\prime \cdot (\vec{r}_i - \vec{r}_j) &= \bigg(\frac{\vec{G}}{2 \vec{k}\cdot\hat{G}} - \frac{\vec{G}}{k}\bigg)\cdot(\vec{r}_i - \vec{r}_j) = 0
\end{align}
where we take $\hat{k} = (\cos\phi\sin\theta, \sin\phi\sin\theta,\cos\theta)$, solving this equation for $\theta$ in the $hkl = 400$ case gives
\begin{equation}
   \theta = \cot ^{-1}\left(\frac{n_x \cos (\phi )-n_y \sin (\phi )}{n_z}\right)+\pi  c_1
\end{equation}
for $n_x, n_y, n_z, c_1 \in \mathbb{Z}$. This defines a family of lattice points that remain in the absorption sum $I$ even in the limit $\lambda \to 0$, resulting a lower bound on $I$ as shown for some example choices of $\hat{k}$ in Fig.~\ref{fig:abs_factor}. This effect is similar in nature to the Laue-case diffraction enhancements where the photoconversion occurs down the scattering planes, minimizing the absorption, as studied in ref.~\cite{Yamaji:2017pep}.

\begin{figure}
    \centering
    \includegraphics[width=0.48\textwidth]{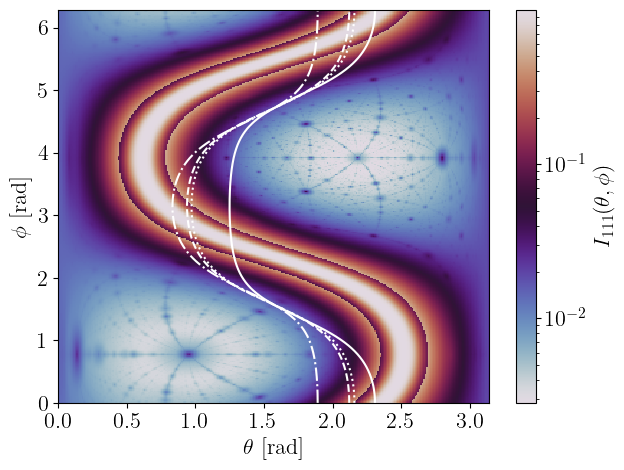}
    \caption{The absorption factor $I(\vec{k}, \vec{G})$ for the $hkl=111$ plane as a function of the incoming ALP direction $\hat{k}(\theta,\phi)$ when the Bragg condition is satisfied and the mean free path is given by the Borrmann anomalous absorption coefficient. Here we take a crystal of cubic volume with side length 5 cm.  The solid white lines trace sample paths of the daily solar angle in January (solid), March (dashed), June (dash-dotted), and September (dotted) at the latitude and longitude of the Gran Sasso site.}
    \label{fig:abs_factor2d}
\end{figure}

In Fig.~\ref{fig:abs_factor2d} the absorption factor $I$ is shown for the plane $\vec{G}(1,1,1)$ as a function of azimuthal and polar angles of the incoming axion momentum $\theta, \phi$ under the Bragg condition. This fixes $k = E_\gamma$ for a given $(\theta, \phi)$, and therefore the attenuation length $\lambda$ given by Eq.~\ref{eq:anomalous_depth}. We see a two prominent features of mitigated absorption in the $S$-shaped band (tracing out a great circle on the 2-sphere), where (i) $I\to 1$ as these $(\theta,\phi)$ combinations correspond to larger energies where the photon absorption cross section falls off as we move further into the $S$, and (ii) there is a jump discontinuity in the $S$-band due to an absorption edge in the photoelectric cross section for germanium at around 11 keV.

\section{Event Rates}
\label{sec:event_rates}
\begin{figure*}[th]
    \centering
    \includegraphics[width=0.45\textwidth]{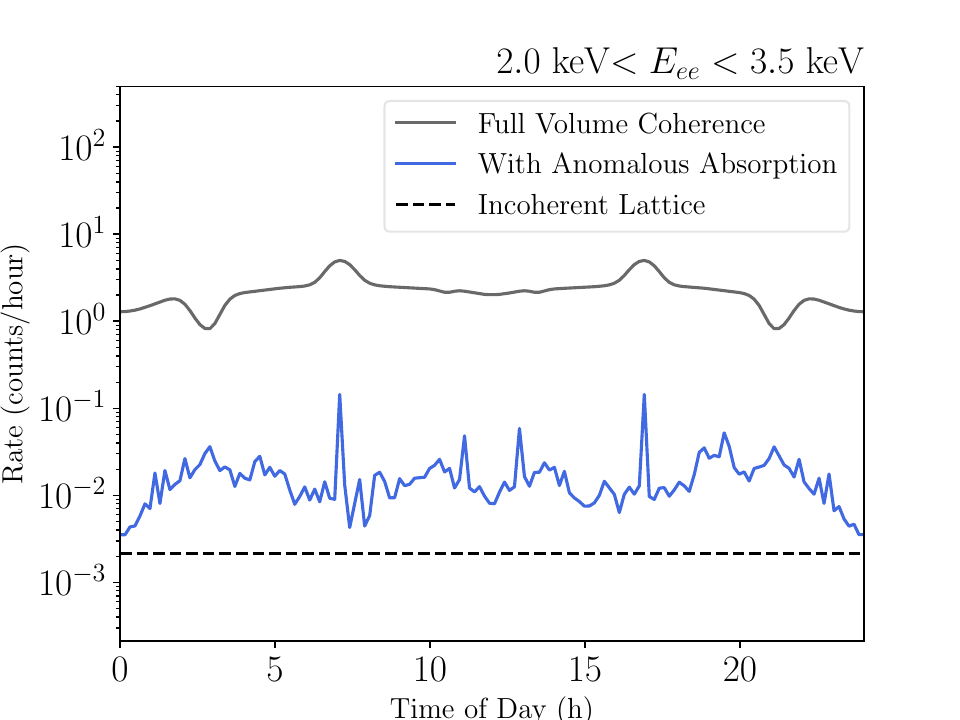}
    \includegraphics[width=0.45\textwidth]{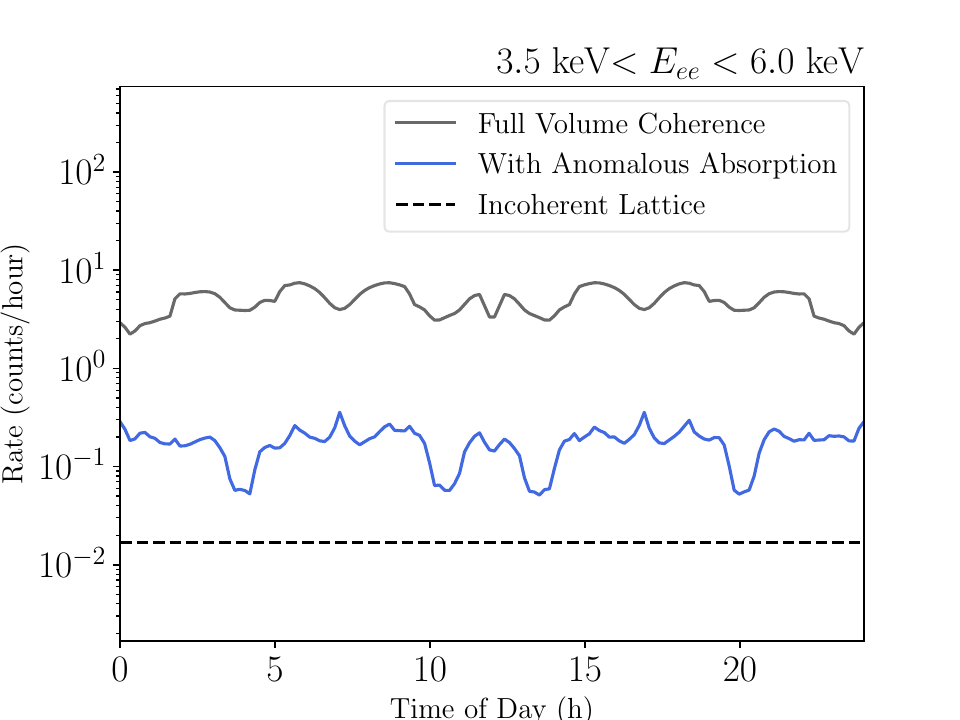}\\
    \includegraphics[width=0.45\textwidth]{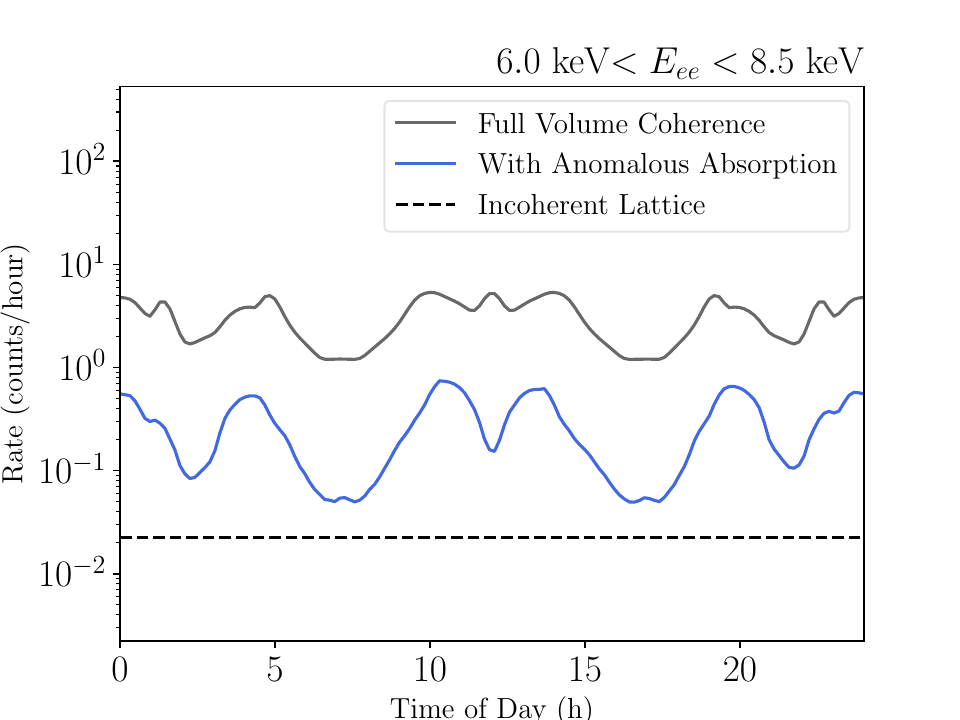}
    \includegraphics[width=0.45\textwidth]{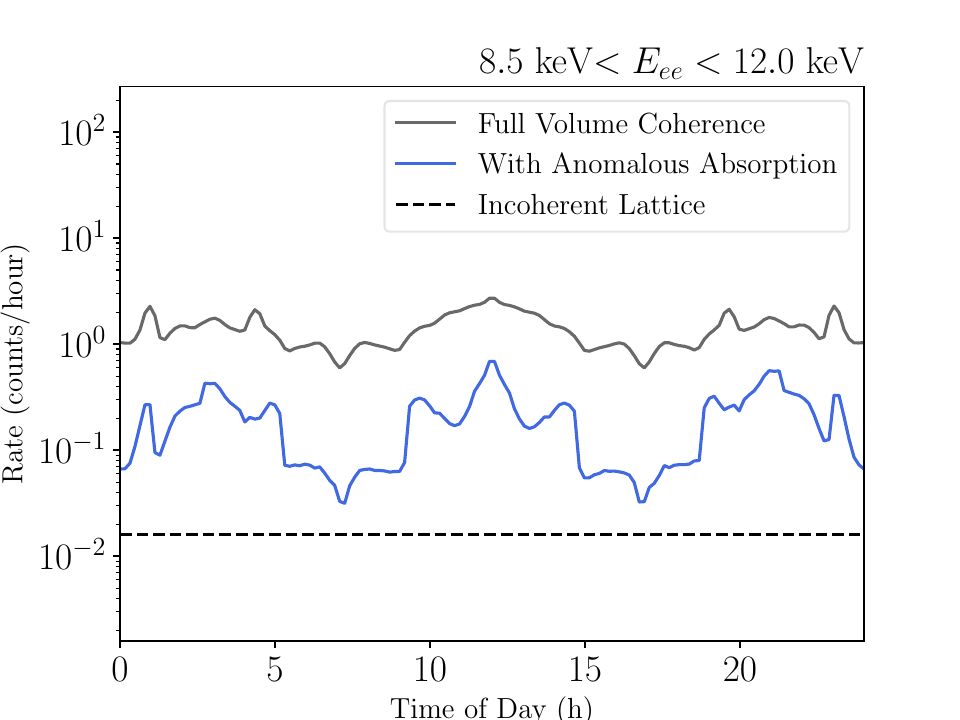}
    \caption{Solar ALP scattering rates in a 250 cm$^3$ Ge crystal, comparing the rates with full volume coherence to ours with anomalous absorption effects included through the absorption factor $I(\vec{k},\vec{G})$. Here we fix the coupling $g_{a\gamma} = 10^{-8}$ GeV$^{-1}$, energy resolution to be $\Delta = 1.0$ keV (for $E_{ee} < 6$ keV) and $\Delta = 1.5$ keV (for $E_{ee} > 6$ keV).}
    \label{fig:ge_rates}
\end{figure*}

The event rate for Primakoff coherent scattering with a perfect crystal worked out in~\cite{Cebrian:1998mu,Bernabei:2001ny,Li:2015tsa} where full-volume coherence was assumed and there is no dependence on the attenuation length\footnote{Notice the factor of $(\hbar c)^3$ rather than $\hbar c$ as written in ref.~\cite{Cebrian:1998mu} for dimensional consistency.}; the event rate in an energy window $[E_1, E_2]$ is
\begin{equation}
\label{eq:event_rate_full_coherency}
    \dfrac{dN}{dt} = \pi g_{a\gamma}^2 (\hbar c)^3 \dfrac{V}{v_\textrm{cell}^2} \sum_{\vec{G}} \bigg[\dfrac{d\Phi_a}{dE_a} \frac{|F_j (\vec{G}) S_j(\vec{G}) |^2}{|\vec{G}|^2} \sin^2 (2\theta) \mathcal{W} \bigg]
\end{equation}
where $S_j$ is the crystal structure factor (see appendix), $F_j$ is the atomic form factor for species $j$, and $d\Phi_a / dE_a$ is the solar axion flux from Primakoff scattering and photon coalescence in the sun~\cite{Redondo:2013wwa, PhysRevD.18.1829}. For the solar axion flux, we take the parameterized form appearing in ref.~\cite{DiLella:2000dn} which expands upon the form originally given by CAST~\cite{CAST:2007jps} by accounting for the axion mass; see Eq.~\ref{eq:solar_fluxes}. The event rate in Eq.~\ref{eq:event_rate_full_coherency} encodes the effect of detector energy resolution $\Delta$ within the function $\mathcal{W}$;
\begin{equation}
    \mathcal{W}(E_a, E_1, E_2, \Delta) = \frac{1}{2}\bigg(\textrm{erf}\bigg(\frac{E_a - E_1}{\sqrt{2}\Delta}\bigg) - \textrm{erf}\bigg(\frac{E_a - E_2}{\sqrt{2}\Delta}\bigg) \bigg)
\end{equation}
The sum over the reciprocal lattice vectors $\vec{G}$ effectively counts the contributions to the coherent scattering from each set of lattice planes, illustrated in Fig.~\ref{fig:csi_planes}. The reader may refer to appendix~\ref{app:crystals} for a compact description of the reciprocal lattice.

At this stage the effect of absorption will simply modify the event rate, as seen in the previous section, by replacing the full coherent volume $V \to V \times I(\vec{k},\vec{G})$ with $\lambda = [\mu_0 (1 - \epsilon(\vec{G}))]^{-1}$, giving
\begin{align}
    \dfrac{dN}{dt} = \pi g_{a\gamma}^2 (\hbar c)^3 \dfrac{V}{v_\textrm{cell}^2} &\sum_{\vec{G}} \bigg[\dfrac{d\Phi_a}{dE_a} \cdot \frac{I(\vec{k},\vec{G})}{|\vec{G}|^2} \nonumber \\
    & \times |F_j (\vec{G}) S_j(\vec{G}) |^2 \sin^2 (2\theta) \mathcal{W} \bigg]
\end{align}
With $\sin^2 (2\theta)$ simplifying to $4(\hat{G}\cdot\hat{k})^2 (1 - (\hat{G}\cdot\hat{k})^2)$~\cite{Bernabei:2001ny} where $\hat{k}$ is the unit vector pointing toward the Sun's location, we have
\begin{align}
    \dfrac{dN}{dt} = \pi g_{a\gamma}^2 (\hbar c)^3 \dfrac{V}{v_\textrm{cell}^2} &\sum_{\vec{G}} I(\vec{k},\vec{G}) \bigg[ \dfrac{d\Phi_a}{dE_a}  |F_j (\vec{G}) S_j(\vec{G}) |^2 \nonumber \\
    & \times \frac{4(\hat{G}\cdot\hat{k})^2 (1 - (\hat{G}\cdot\hat{k})^2)}{|\vec{G}|^2} \mathcal{W} \bigg]
\end{align}
At this stage, we have also used the Bragg condition $E_a = \hbar c|\vec{G}|^2 / (2 \hat{k}\cdot\vec{G})$. The time dependence is encoded in the solar position, which we can express through $\hat{k} = (\cos\phi \sin\theta, \sin\phi \sin\theta, \cos\theta)$ for $\theta = \theta(t)$ and $\phi = \phi(t)$. For the solar angle as a function of time and geolocation, we use the NREL solar position algorithm~\cite{nrel}. 

In principle, the sum over reciprocal lattice vectors $\vec{G}$ is taken to arbitrarily large combinations $(h,k,l)$, but due to the $1/|\vec{G}|^2$ suppression and the upper limit of the solar axion flux of around $\sim 20$ keV, we can safely truncate the sum at $max\{h,k,l\}=5$.

The corresponding event rates for various energy windows are shown in Fig.~\ref{fig:ge_rates} for Ge crystal, where we compare the relative enhancements with and without the Borrmann effect to the case of full-volume coherence and to the case of incoherent scattering on an amorphous lattice\footnote{Atomic Primakoff scattering is still coherent here; we only turn off the coherence at the level of the lattice for the sake of comparison with scattering on amorphous materials, in this case, amorphous germanium.}. The fluctuating features in the event rate are the result of the sum over $\vec{G}$ which contributes to the Bragg peaks. Here we have assumed a volume of 260 cm$^3$ (corresponding roughly to the volumetric size of a SuperCDMS germanium module), and so the relative suppression for each $\vec{G}$ lattice plane goes like $V^{1/3} / \lambda(\vec{k},\vec{G})$, giving a suppression on the order of $10^2$ compared to the full-volume coherence assumption. 

\begin{figure}[h]
    \centering
    \includegraphics[width=0.5\textwidth]{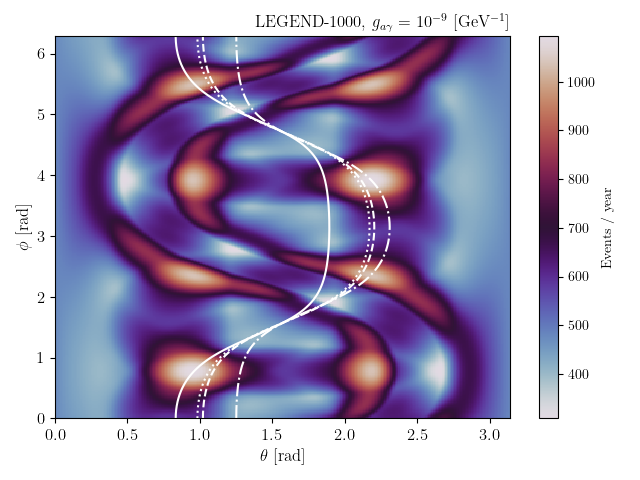}
    \caption{Event rates in germanium as a function of incident angles $\theta, \phi$ for the integrated energy range $(1, 10)$ keV and for reciprocal lattice planes $(h,k,l)$ up to $max\{h,k,l\} = 5$. The solid white lines trace sample paths of the daily solar angle in January (solid), March (dashed), June (dash-dotted), and September (dotted) at the latitude and longitude of the Gran Sasso site.}
    \label{fig:2d_rates}
\end{figure}

The time-dependence can be visualized further by viewing the event rates as a function of incident angles integrated across the whole solar axion energy window, as shown in Fig.~\ref{fig:2d_rates}. Depending on the time of year, different sets of Bragg peaks will be traced over during the day, inducing an annual modulation in addition to the intra-day modulation of the signal.

\begin{figure}
    \centering
    \includegraphics[width=0.5\textwidth]{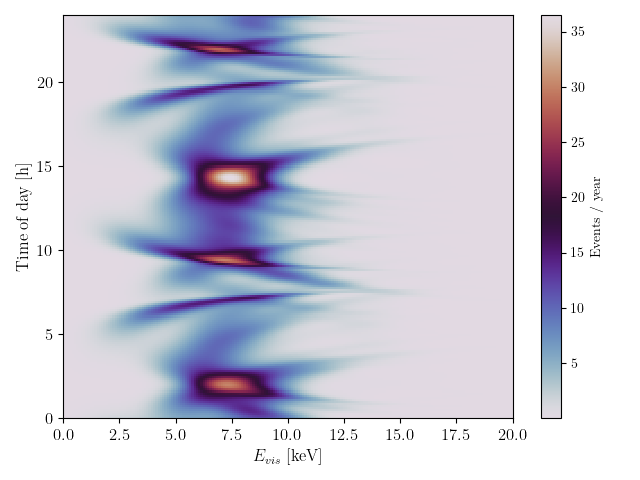}
    \caption{Differential energy-time event rate with energy resolution $\Delta = 2.5$ keV. The time of year was taken to be January at the latitude and longitude of the Gran Sasso site.}
    \label{fig:2d_rates_E_t}
\end{figure}

Since the time of day fixes the solar zenith and azimuth $(\theta, \phi)$, we can finally show the spectrum of the Primakoff signal as a function of energy deposition and time of day; see Fig.~\ref{fig:2d_rates_E_t}.

\section{Projected Sensitivities for Solar Axion Searches}
\label{sec:sensitivities}
\begin{figure}[!hbt]
    \centering
    \includegraphics[width=0.5\textwidth]{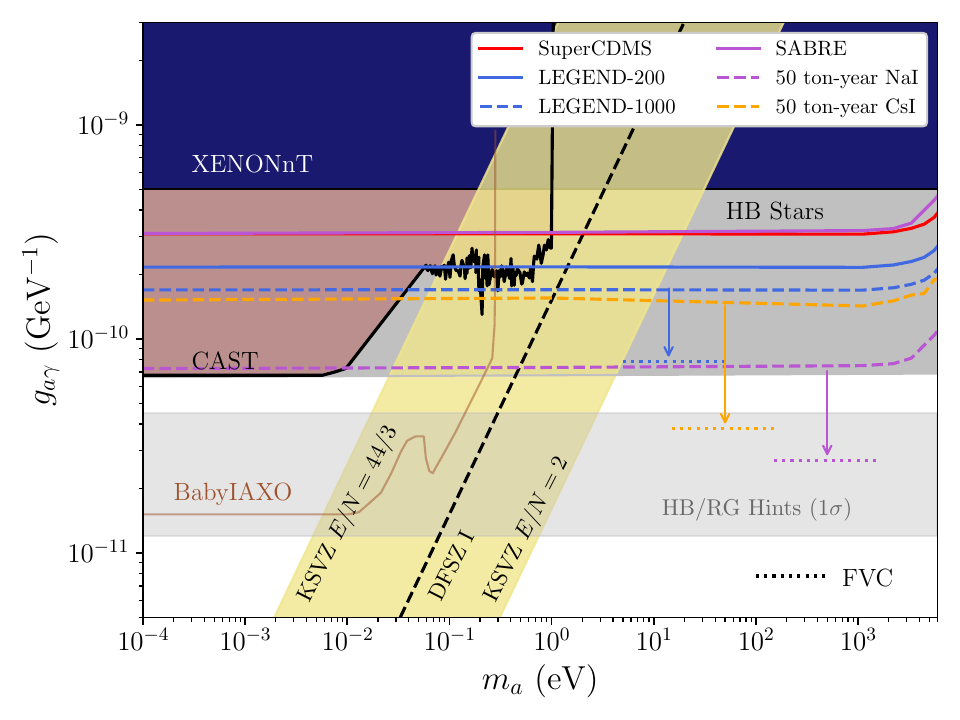}
    \caption{We show sensitivity projections for Ge experiments (SuperCDMS, LEGEND-200, and LEGEND-1000), NaI (SABRE) and multi-tonne CsI and NaI setups with absorption effects included. Existing bounds from solar axion searches at XENONnT~\cite{XENON:2022mpc} and CAST~\cite{CAST:2007jps} as well as the constraint from HB star cooling~\cite{Ayala:2014pea} are shown. The complementary reach of axion helioscope BabyIAXO~\cite{Armengaud:2019uso} is shown for comparison. Changes to the projected reach for crystal detectors if full volume coherence (FVC) is restored are indicated by the dotted lines.}
    \label{fig:sensitivity}
\end{figure}

\begin{table*}[ht!]
    \centering
    \begin{tabular}{|l|l|c|c|c|c|}
    \hline
         Experiment & \thead{Module Mass\\$\times$no. Modules} & \thead{Total Mass} & \thead{Energy\\Resolution} & Threshold & Exposure (ton-years) \\
         \hline
         SuperCDMS (Ge) & 1.4 kg $\times$ 18 & 25.2 kg & 2.5 keV & 1 keV & 0.1 \\
         SuperCDMS (Si) & 0.6 kg $\times$ 6 & 3.6 kg & 2.5 keV & 1 keV & 0.0144 \\
         LEGEND-200 (Ge) & 2.6 kg $\times$ 75 & 195 kg & 2.5 keV & 1 keV & 0.78 \\
         LEGEND-1000 (Ge) & 2.6 kg $\times$ 400 & 1 tonne & 2.5 keV & 1 keV & 5 \\
         SABRE (NaI) & 2 kg $\times$ 25 & 50 kg & 1 keV & 1 keV & 0.15\\
         10-tonne NaI & 2 kg $\times$ 2500 & 5 tonne & 1 keV & 1 keV & 50 \\
         10-tonne CsI & 2 kg $\times$ 2500 & 5 tonne & 1 keV & 1 keV & 50 \\
         \hline
    \end{tabular}
    \caption{Assumed detector parameters for the SuperCDMS~\cite{SuperCDMS:2022kse}, LEGEND~\cite{LEGEND:2021bnm}, and SABRE~\cite{SABRE:2018lfp} configurations.}
    \label{tab:detectors}
\end{table*}
We forecast the event rates for SuperCDMS~\cite{SuperCDMS:2022kse}, LEGEND-200, LEGEND-1000, SABRE, in addition to envisioned multi-tonne setups, with detector specifications listed in Table~\ref{tab:detectors}. For the background-free limits, we look for the Poisson 90\% CL corresponding to $\simeq 3$ events observed for a given exposure. The projected reach over the $(g_{a\gamma} - m_a)$ parameter space for these detector benchmarks is shown in Fig.~\ref{fig:sensitivity}, where we show projections including the effects of absorption and the Borrmann enhancement to the absorption length, in addition to the projected limits assuming full volume coherence (FVC), i.e. $I(\vec{k},\vec{G})\to1$, indicated by the arrows and dotted lines. 

The QCD axion parameter space is shown (yellow band) for the Kim-Shifman-Vainshtein-Zakharov (KSVZ) type~\cite{PhysRevLett.43.103, SHIFMAN1980493} and Dine-Fischler-Srednicki-Zhitnitsky (DFSZ) type benchmark models~\cite{Zhitnitsky:1980tq,DINE1981199,Dine:1981rt,Dine:1982ah}, where the range is defined by taking the anomaly ratios of $E/N = 44/3$ to $E/N = 2$~\cite{DiLuzio:2020wdo}, although the space of heavier masses is also possible in high-quality axion models and other scenarios~\cite{Kivel:2022emq, Valenti:2022tsc}. To probe this model parameter space beyond the existing bounds from CAST and horizontal branch (HB) stars, when FVC is maintained, multi-tonne scale experiments are needed. Additionally, the stellar cooling hints that could be explained by ALPs with $g_{a\gamma}\lesssim 10^{-11}$ GeV$^{-1}$ (and for non-vanishing $g_{ae} \simeq 10^{-13}$), are also shown in Fig.~\ref{fig:sensitivity}, indicated by the gray band (1$\sigma$) and down to vanishing $g_{a\gamma}$~\cite{Giannotti:2015kwo,Hoof:2018ieb,Ayala:2014pea}. These hints, though mild, could be tested by the multi-tonne setups with FVC restored.

With the effects of absorption included, we project SuperCDMS, LEGEND, and SABRE to test parameter space unexplored by laboratory-based probes beyond the CAST and XENONnT constraints for $m_a \gtrsim 1$ eV, but already excluded by HB stars constraints. However, multi-tonne CsI and NaI setups would extend this to nearly cover the HB stars exclusion. Similar reach could in principle be found when considering the joint parameter space of multiple ALP couplings to photons, electrons, and nucleons~\cite{Dent:2020jhf}. For instance, by considering the $^{57}$Fe solar axion flux, one could look for 14.4 keV energy signatures and their Bragg-Primakoff peaks, although the sensitivity would likely contend with astrophysics constraints as well~\cite{Hardy:2016kme}.

The existing bounds from DAMA~\cite{Bernabei:2004fi}, CUORE~\cite{Li:2015tyq}, Edelweiss-II~\cite{Armengaud_2013}, SOLAX~\cite{PhysRevLett.81.5068}, COSME~\cite{COSME:2001jci}, CDMS~\cite{CDMS:2009fba}, and Majorana~\cite{Majorana:2022bse} are not shown here, but their exclusions would necessarily shift to larger coupling values to account for absorption effects in the Bragg-Primakoff rates, depending on the detector volume and material. Note that the relative reach between NaI and CsI crystals is relatively suppressed when absorption is included here, due to the behavior of the imaginary form factor for CsI giving more modest Borrmann enhancements at the lower reciprocal lattice planes; see Fig.~\ref{fig:borrmann_by_mat}. In order to push the sensitivity envelope beyond the current bounds by CAST and HB stars, even with multi-tonne setups, the absorption effects need to be mitigated. Some possibilities are discussed in the next section.

\section{Restoring Coherence}
\label{sec:restoring_coh}
There may be ways to recover the sensitivity initially projected in the case of full-volume coherence by mitigating the loss of coherence due to absorption. These are of course speculative routes. Some of these routes for future work are enumerated below;
\begin{enumerate}
    \item Since the attenuation of the coherent volume is direction-dependent, as shown in Fig.~\ref{fig:abs_factor2d}, one could imagine optimizing a detector geometry such that the size and orientation relative to the incoming flux of axions is ideal, maximizing use of the Laue-type scattering and Borrmann effect to minimize the absorption. This would require precise knowledge of the crystal purity and plane orientation obtained from X-ray measurements.
    \item Along a similar vein, since the effects of absorption are minimized when the detector scale $V^{1/3}$ becomes comparable to the photon mean free path $\lambda$, one could instead prefer to use smaller detector volumes but with a large total mass partitioned into many individual modules. As long as each module is optically insulated from the others, the loss of coherence due to absorption will be contained within each module and the suppression to the event rate can be mitigated.
    \item It might be possible to apply the principles in this work to radioisotope experiments like those proposed in ref.~\cite{PhysRevD.99.035025,Dent:2021jnf}, where a keV-scale nuclear transition line (e.g. the 14.4 keV line of $^{57}$Fe) could source ALPs through a coupling to nucleons. Subsequent detection by an array of crystals encasing the radioactive source searching for transition photons of known energy Primakoff-converting in the crystal would leave a missing energy signature in the detector. By looking for disappearing keV-scale transitions the signal rate would enjoy the coherent enhancement relative to the incoherent scattering considered in ref.~\cite{Dent:2021jnf}.
    \item A dedicated keV photon source that would impinge on a crystal detector could fire at a fixed angle of incidence such that the event rate enhancement from the Borrmann effect and Laue effects are optimized and full volume coherence is restored as best as possible. One might achieve this with a keV laser~\cite{lanlLaser} or synchrotron sources in a similar fashion to LSW experiments~\cite{Yamaji:2018ufo, PhysRevLett.105.250405, PhysRevLett.118.071803, PhysRevD.105.035031,Bahre:2013ywa}. By performing a similar ``missing'' photon search as the one discussed above, the event rate for the detection of missing energy will be proportional to $g_{a\gamma}^2$, rather than $g_{a\gamma}^4$ as in solar axion searches, greatly enhancing the sensitivity.\\
\end{enumerate}

In the case where we assume full volume coherence, shown in Fig.~\ref{fig:sensitivity}, dotted lines, ton-scale setups like LEGEND-200 and LEGEND-1000 can reach significantly smaller couplings, probing values of $g_{a\gamma}$ beyond the existing bounds fom HB Stars~\cite{Ayala:2014pea,Giannotti:2015kwo} and CAST~\cite{CAST:2017uph} for masses $m_a \lesssim 10$ keV, losing sensitivity for higher masses for which the axion production rates from photon coalescence and Primakoff scattering are diminished (see also Fig.~\ref{fig:solar_fluxes}). These reach more than an order of magnitude lower in the coupling than previous Bragg-Primakoff solar axion searches. 

\section{Conclusions}
\label{sec:conclude}
In this work, we have taken into account a more proper estimate of the effects of anomalous absorption into the event rate, i.e. via the Borrmann effect on the coherence condition of Bragg-Primakoff photoconversion of solar axions. The sensitivity of crystal technologies used in the SuperCDMS, LEGEND, and SABRE setups has been demonstrated, and we find that the inclusion of absorption effects even with Borrmann-enhanced signal rates still would require multi-tonne scale detectors to surpass the existing astrophysical constraints in sensitivity to ALPs. However, a dedicated study with a thorough and careful treatment of the absorption suppression and Borrmann effects is definitely needed to better understand its impact on experiments that utilize Bragg-Primakoff conversion. In particular, the evaluation of the imaginary form factor in other crystals (namely, PbWO$_4$ may be an interesting option) would help determine potential enhancements to the anomalous absorption effect in other detector materials.

Crystal detector technologies are also necessary tools to discriminate axion-like particle signals from other types of BSM and neutrino signatures, with high sensitivity to time modulation from the directional sensitivity of Bragg-Primakoff scattering. This is a powerful tool for background rejection as well, and ideally a joint analysis of multiple detectors situated at different latitudes and longitudes would benefit greatly from leveraging the time modulation of the signal. They are also complimentary to future helioscope experiments like IAXO; while the projected reach for IAXO over the axion-photon coupling parameter space is vast, the sensitivity to solar axions with masses $m_a \gtrsim 1$ eV becomes weaker to coherent Primakoff conversion in magnetic field helioscopes. Sensitivity to this region of parameter space is necessary in order to test QCD axions, especially in non-traditional models of high quality axions and the like, which have parametrically larger masses~\cite{Gaillard:2018xgk, Kivel:2022emq, Hook:2019qoh}. It was shown in ref.~\cite{Dent:2020jhf} that future liquid noble gas detectors for dark matter direct detection at kiloton-year scales could begin to probe couplings beyond the astrophysics constraints for axion-like particles, while in this work we find that equivalent reach is possible at ton-year exposures with crystal detector technology, if utilized to its fullest potential. The presence of complimentary searches at these mass scales is essential for a complete test of the axion solution to the strong CP problem and the broader space of ALPs.

\section*{Acknowledgements}
We are very grateful to Imran Alkhatib, Miriam Diamond, Amirata Sattari Javid, and John Sipe for the vigorous discussions and studies on the theoretical treatment of coherent Primakoff scattering in crystals and the comparison of numerical computations. We graciously thank Tomohiro Yamaji for the insight on Laue-type diffraction, Timon Emken for the technical correspondence on the \texttt{DarkART} package, and Alexander Poddubny for the useful comments on Biagini's theory of anomalous absorption. The work of BD and AT is supported by the DOE Grant No. DE-SC0010813. JBD acknowledges support from the National Science Foundation under grant no. PHY-2112799. Portions of this research were conducted with the advanced computing resources provided by Texas A\&M High Performance Research Computing. We also thank the Center for Theoretical Underground Physics and Related Areas (CETUP*) and SURF for facilitating portions of this research.

\bibliography{main}

\begin{thebibliography}{82}%
\makeatletter
\providecommand \@ifxundefined [1]{%
 \@ifx{#1\undefined}
}%
\providecommand \@ifnum [1]{%
 \ifnum #1\expandafter \@firstoftwo
 \else \expandafter \@secondoftwo
 \fi
}%
\providecommand \@ifx [1]{%
 \ifx #1\expandafter \@firstoftwo
 \else \expandafter \@secondoftwo
 \fi
}%
\providecommand \natexlab [1]{#1}%
\providecommand \enquote  [1]{``#1''}%
\providecommand \bibnamefont  [1]{#1}%
\providecommand \bibfnamefont [1]{#1}%
\providecommand \citenamefont [1]{#1}%
\providecommand \href@noop [0]{\@secondoftwo}%
\providecommand \href [0]{\begingroup \@sanitize@url \@href}%
\providecommand \@href[1]{\@@startlink{#1}\@@href}%
\providecommand \@@href[1]{\endgroup#1\@@endlink}%
\providecommand \@sanitize@url [0]{\catcode `\\12\catcode `\$12\catcode
  `\&12\catcode `\#12\catcode `\^12\catcode `\_12\catcode `\%12\relax}%
\providecommand \@@startlink[1]{}%
\providecommand \@@endlink[0]{}%
\providecommand \url  [0]{\begingroup\@sanitize@url \@url }%
\providecommand \@url [1]{\endgroup\@href {#1}{\urlprefix }}%
\providecommand \urlprefix  [0]{URL }%
\providecommand \Eprint [0]{\href }%
\providecommand \doibase [0]{http://dx.doi.org/}%
\providecommand \selectlanguage [0]{\@gobble}%
\providecommand \bibinfo  [0]{\@secondoftwo}%
\providecommand \bibfield  [0]{\@secondoftwo}%
\providecommand \translation [1]{[#1]}%
\providecommand \BibitemOpen [0]{}%
\providecommand \bibitemStop [0]{}%
\providecommand \bibitemNoStop [0]{.\EOS\space}%
\providecommand \EOS [0]{\spacefactor3000\relax}%
\providecommand \BibitemShut  [1]{\csname bibitem#1\endcsname}%
\let\auto@bib@innerbib\@empty
\bibitem [{\citenamefont {Peccei}\ and\ \citenamefont
  {Quinn}(1977)}]{Peccei:1977hh}%
  \BibitemOpen
  \bibfield  {author} {\bibinfo {author} {\bibfnamefont {R.~D.}\ \bibnamefont
  {Peccei}}\ and\ \bibinfo {author} {\bibfnamefont {H.~R.}\ \bibnamefont
  {Quinn}},\ }\href {\doibase 10.1103/PhysRevLett.38.1440} {\bibfield
  {journal} {\bibinfo  {journal} {Phys. Rev. Lett.}\ }\textbf {\bibinfo
  {volume} {38}},\ \bibinfo {pages} {1440} (\bibinfo {year}
  {1977})}\BibitemShut {NoStop}%
\bibitem [{\citenamefont {Wilczek}(1978)}]{Wilczek:1977pj}%
  \BibitemOpen
  \bibfield  {author} {\bibinfo {author} {\bibfnamefont {F.}~\bibnamefont
  {Wilczek}},\ }\href {\doibase 10.1103/PhysRevLett.40.279} {\bibfield
  {journal} {\bibinfo  {journal} {Phys. Rev. Lett.}\ }\textbf {\bibinfo
  {volume} {40}},\ \bibinfo {pages} {279} (\bibinfo {year} {1978})}\BibitemShut
  {NoStop}%
\bibitem [{\citenamefont {Weinberg}(1978)}]{Weinberg:1977ma}%
  \BibitemOpen
  \bibfield  {author} {\bibinfo {author} {\bibfnamefont {S.}~\bibnamefont
  {Weinberg}},\ }\href {\doibase 10.1103/PhysRevLett.40.223} {\bibfield
  {journal} {\bibinfo  {journal} {Phys. Rev. Lett.}\ }\textbf {\bibinfo
  {volume} {40}},\ \bibinfo {pages} {223} (\bibinfo {year} {1978})}\BibitemShut
  {NoStop}%
\bibitem [{\citenamefont {Preskill}\ \emph {et~al.}(1983)\citenamefont
  {Preskill}, \citenamefont {Wise},\ and\ \citenamefont
  {Wilczek}}]{Preskill:1982cy}%
  \BibitemOpen
  \bibfield  {author} {\bibinfo {author} {\bibfnamefont {J.}~\bibnamefont
  {Preskill}}, \bibinfo {author} {\bibfnamefont {M.~B.}\ \bibnamefont {Wise}},
  \ and\ \bibinfo {author} {\bibfnamefont {F.}~\bibnamefont {Wilczek}},\ }\href
  {\doibase 10.1016/0370-2693(83)90637-8} {\bibfield  {journal} {\bibinfo
  {journal} {Phys. Lett. B}\ }\textbf {\bibinfo {volume} {120}},\ \bibinfo
  {pages} {127} (\bibinfo {year} {1983})}\BibitemShut {NoStop}%
\bibitem [{\citenamefont {Abbott}\ and\ \citenamefont
  {Sikivie}(1983)}]{Abbott:1982af}%
  \BibitemOpen
  \bibfield  {author} {\bibinfo {author} {\bibfnamefont {L.}~\bibnamefont
  {Abbott}}\ and\ \bibinfo {author} {\bibfnamefont {P.}~\bibnamefont
  {Sikivie}},\ }\href {\doibase 10.1016/0370-2693(83)90638-X} {\bibfield
  {journal} {\bibinfo  {journal} {Phys. Lett. B}\ }\textbf {\bibinfo {volume}
  {120}},\ \bibinfo {pages} {133} (\bibinfo {year} {1983})}\BibitemShut
  {NoStop}%
\bibitem [{\citenamefont {Dine}\ and\ \citenamefont
  {Fischler}(1983)}]{Dine:1982ah}%
  \BibitemOpen
  \bibfield  {author} {\bibinfo {author} {\bibfnamefont {M.}~\bibnamefont
  {Dine}}\ and\ \bibinfo {author} {\bibfnamefont {W.}~\bibnamefont
  {Fischler}},\ }\href {\doibase 10.1016/0370-2693(83)90639-1} {\bibfield
  {journal} {\bibinfo  {journal} {Phys. Lett. B}\ }\textbf {\bibinfo {volume}
  {120}},\ \bibinfo {pages} {137} (\bibinfo {year} {1983})}\BibitemShut
  {NoStop}%
\bibitem [{\citenamefont {Battaglieri}\ \emph {et~al.}(2017)\citenamefont
  {Battaglieri} \emph {et~al.}}]{Battaglieri:2017aum}%
  \BibitemOpen
  \bibfield  {author} {\bibinfo {author} {\bibfnamefont {M.}~\bibnamefont
  {Battaglieri}} \emph {et~al.},\ }in\ \href
  {http://lss.fnal.gov/archive/2017/conf/fermilab-conf-17-282-ae-ppd-t.pdf}
  {\emph {\bibinfo {booktitle} {{U.S. Cosmic Visions: New Ideas in Dark Matter
  College Park, MD, USA, March 23-25, 2017}}}}\ (\bibinfo {year} {2017})\
  \Eprint {http://arxiv.org/abs/1707.04591} {arXiv:1707.04591 [hep-ph]}
  \BibitemShut {NoStop}%
\bibitem [{\citenamefont {Marsh}(2016)}]{Marsh:2015xka}%
  \BibitemOpen
  \bibfield  {author} {\bibinfo {author} {\bibfnamefont {D.~J.~E.}\
  \bibnamefont {Marsh}},\ }\href {\doibase 10.1016/j.physrep.2016.06.005}
  {\bibfield  {journal} {\bibinfo  {journal} {Phys. Rept.}\ }\textbf {\bibinfo
  {volume} {643}},\ \bibinfo {pages} {1} (\bibinfo {year} {2016})},\ \Eprint
  {http://arxiv.org/abs/1510.07633} {arXiv:1510.07633 [astro-ph.CO]}
  \BibitemShut {NoStop}%
\bibitem [{\citenamefont {Arias}\ \emph {et~al.}(2012)\citenamefont {Arias},
  \citenamefont {Cadamuro}, \citenamefont {Goodsell}, \citenamefont {Jaeckel},
  \citenamefont {Redondo},\ and\ \citenamefont {Ringwald}}]{Arias:2012az}%
  \BibitemOpen
  \bibfield  {author} {\bibinfo {author} {\bibfnamefont {P.}~\bibnamefont
  {Arias}}, \bibinfo {author} {\bibfnamefont {D.}~\bibnamefont {Cadamuro}},
  \bibinfo {author} {\bibfnamefont {M.}~\bibnamefont {Goodsell}}, \bibinfo
  {author} {\bibfnamefont {J.}~\bibnamefont {Jaeckel}}, \bibinfo {author}
  {\bibfnamefont {J.}~\bibnamefont {Redondo}}, \ and\ \bibinfo {author}
  {\bibfnamefont {A.}~\bibnamefont {Ringwald}},\ }\href {\doibase
  10.1088/1475-7516/2012/06/013} {\bibfield  {journal} {\bibinfo  {journal}
  {JCAP}\ }\textbf {\bibinfo {volume} {06}},\ \bibinfo {pages} {013} (\bibinfo
  {year} {2012})},\ \Eprint {http://arxiv.org/abs/1201.5902} {arXiv:1201.5902
  [hep-ph]} \BibitemShut {NoStop}%
\bibitem [{\citenamefont {Duffy}\ and\ \citenamefont {van
  Bibber}(2009)}]{Duffy:2009ig}%
  \BibitemOpen
  \bibfield  {author} {\bibinfo {author} {\bibfnamefont {L.~D.}\ \bibnamefont
  {Duffy}}\ and\ \bibinfo {author} {\bibfnamefont {K.}~\bibnamefont {van
  Bibber}},\ }\href {\doibase 10.1088/1367-2630/11/10/105008} {\bibfield
  {journal} {\bibinfo  {journal} {New J. Phys.}\ }\textbf {\bibinfo {volume}
  {11}},\ \bibinfo {pages} {105008} (\bibinfo {year} {2009})},\ \Eprint
  {http://arxiv.org/abs/0904.3346} {arXiv:0904.3346 [hep-ph]} \BibitemShut
  {NoStop}%
\bibitem [{\citenamefont {Adams}\ \emph {et~al.}(2022)\citenamefont {Adams}
  \emph {et~al.}}]{Adams:2022pbo}%
  \BibitemOpen
  \bibfield  {author} {\bibinfo {author} {\bibfnamefont {C.~B.}\ \bibnamefont
  {Adams}} \emph {et~al.},\ }in\ \href@noop {} {\emph {\bibinfo {booktitle}
  {{2022 Snowmass Summer Study}}}}\ (\bibinfo {year} {2022})\ \Eprint
  {http://arxiv.org/abs/2203.14923} {arXiv:2203.14923 [hep-ex]} \BibitemShut
  {NoStop}%
\bibitem [{\citenamefont {Chikashige}\ \emph {et~al.}(1981)\citenamefont
  {Chikashige}, \citenamefont {Mohapatra},\ and\ \citenamefont
  {Peccei}}]{Chikashige:1980ui}%
  \BibitemOpen
  \bibfield  {author} {\bibinfo {author} {\bibfnamefont {Y.}~\bibnamefont
  {Chikashige}}, \bibinfo {author} {\bibfnamefont {R.~N.}\ \bibnamefont
  {Mohapatra}}, \ and\ \bibinfo {author} {\bibfnamefont {R.~D.}\ \bibnamefont
  {Peccei}},\ }\href {\doibase 10.1016/0370-2693(81)90011-3} {\bibfield
  {journal} {\bibinfo  {journal} {Phys. Lett. B}\ }\textbf {\bibinfo {volume}
  {98}},\ \bibinfo {pages} {265} (\bibinfo {year} {1981})}\BibitemShut
  {NoStop}%
\bibitem [{\citenamefont {Svrcek}\ and\ \citenamefont
  {Witten}(2006)}]{Svrcek:2006yi}%
  \BibitemOpen
  \bibfield  {author} {\bibinfo {author} {\bibfnamefont {P.}~\bibnamefont
  {Svrcek}}\ and\ \bibinfo {author} {\bibfnamefont {E.}~\bibnamefont
  {Witten}},\ }\href {\doibase 10.1088/1126-6708/2006/06/051} {\bibfield
  {journal} {\bibinfo  {journal} {JHEP}\ }\textbf {\bibinfo {volume} {06}},\
  \bibinfo {pages} {051} (\bibinfo {year} {2006})},\ \Eprint
  {http://arxiv.org/abs/hep-th/0605206} {arXiv:hep-th/0605206} \BibitemShut
  {NoStop}%
\bibitem [{\citenamefont {Goodsell}\ and\ \citenamefont
  {Ringwald}(2010)}]{Goodsell:2010ie}%
  \BibitemOpen
  \bibfield  {author} {\bibinfo {author} {\bibfnamefont {M.}~\bibnamefont
  {Goodsell}}\ and\ \bibinfo {author} {\bibfnamefont {A.}~\bibnamefont
  {Ringwald}},\ }\href {\doibase 10.1002/prop.201000026} {\bibfield  {journal}
  {\bibinfo  {journal} {Fortsch. Phys.}\ }\textbf {\bibinfo {volume} {58}},\
  \bibinfo {pages} {716} (\bibinfo {year} {2010})},\ \Eprint
  {http://arxiv.org/abs/1002.1840} {arXiv:1002.1840 [hep-th]} \BibitemShut
  {NoStop}%
\bibitem [{\citenamefont {Sikivie}(1983)}]{Sikivie:1983ip}%
  \BibitemOpen
  \bibfield  {author} {\bibinfo {author} {\bibfnamefont {P.}~\bibnamefont
  {Sikivie}},\ }\href {\doibase 10.1103/PhysRevLett.51.1415} {\bibfield
  {journal} {\bibinfo  {journal} {Phys. Rev. Lett.}\ }\textbf {\bibinfo
  {volume} {51}},\ \bibinfo {pages} {1415} (\bibinfo {year} {1983})},\ \bibinfo
  {note} {[Erratum: Phys.Rev.Lett. 52, 695 (1984)]}\BibitemShut {NoStop}%
\bibitem [{\citenamefont {Raffelt}(1996)}]{Raffelt:1996wa}%
  \BibitemOpen
  \bibfield  {author} {\bibinfo {author} {\bibfnamefont {G.~G.}\ \bibnamefont
  {Raffelt}},\ }\href@noop {} {\emph {\bibinfo {title} {{Stars as laboratories
  for fundamental physics}: {The astrophysics of neutrinos, axions, and other
  weakly interacting particles}}}}\ (\bibinfo {year} {1996})\BibitemShut
  {NoStop}%
\bibitem [{\citenamefont {Bernabei}\ \emph {et~al.}(2004)\citenamefont
  {Bernabei} \emph {et~al.}}]{Bernabei:2004fi}%
  \BibitemOpen
  \bibfield  {author} {\bibinfo {author} {\bibfnamefont {R.}~\bibnamefont
  {Bernabei}} \emph {et~al.},\ }\href@noop {} {\bibfield  {journal} {\bibinfo
  {journal} {Frascati Phys. Ser.}\ }\textbf {\bibinfo {volume} {37}},\ \bibinfo
  {pages} {211} (\bibinfo {year} {2004})}\BibitemShut {NoStop}%
\bibitem [{\citenamefont {Li}\ \emph {et~al.}(2016)\citenamefont {Li},
  \citenamefont {Creswick}, \citenamefont {Avignone},\ and\ \citenamefont
  {Wang}}]{Li:2015tyq}%
  \BibitemOpen
  \bibfield  {author} {\bibinfo {author} {\bibfnamefont {D.}~\bibnamefont
  {Li}}, \bibinfo {author} {\bibfnamefont {R.~J.}\ \bibnamefont {Creswick}},
  \bibinfo {author} {\bibfnamefont {F.~T.}\ \bibnamefont {Avignone}}, \ and\
  \bibinfo {author} {\bibfnamefont {Y.}~\bibnamefont {Wang}},\ }\href {\doibase
  10.1088/1475-7516/2016/02/031} {\bibfield  {journal} {\bibinfo  {journal}
  {JCAP}\ }\textbf {\bibinfo {volume} {02}},\ \bibinfo {pages} {031} (\bibinfo
  {year} {2016})},\ \Eprint {http://arxiv.org/abs/1512.01298} {arXiv:1512.01298
  [astro-ph.CO]} \BibitemShut {NoStop}%
\bibitem [{\citenamefont {Li}\ \emph {et~al.}(2015)\citenamefont {Li},
  \citenamefont {Creswick}, \citenamefont {Avignone},\ and\ \citenamefont
  {Wang}}]{Li:2015tsa}%
  \BibitemOpen
  \bibfield  {author} {\bibinfo {author} {\bibfnamefont {D.}~\bibnamefont
  {Li}}, \bibinfo {author} {\bibfnamefont {R.~J.}\ \bibnamefont {Creswick}},
  \bibinfo {author} {\bibfnamefont {F.~T.}\ \bibnamefont {Avignone}}, \ and\
  \bibinfo {author} {\bibfnamefont {Y.}~\bibnamefont {Wang}},\ }\href {\doibase
  10.1088/1475-7516/2015/10/065} {\bibfield  {journal} {\bibinfo  {journal}
  {JCAP}\ }\textbf {\bibinfo {volume} {10}},\ \bibinfo {pages} {065} (\bibinfo
  {year} {2015})},\ \Eprint {http://arxiv.org/abs/1507.00603} {arXiv:1507.00603
  [astro-ph.CO]} \BibitemShut {NoStop}%
\bibitem [{\citenamefont {Armengaud}\ \emph {et~al.}(2013)\citenamefont
  {Armengaud} \emph {et~al.}}]{Armengaud_2013}%
  \BibitemOpen
  \bibfield  {author} {\bibinfo {author} {\bibfnamefont {E.}~\bibnamefont
  {Armengaud}} \emph {et~al.},\ }\href {\doibase 10.1088/1475-7516/2013/11/067}
  {\bibfield  {journal} {\bibinfo  {journal} {Journal of Cosmology and
  Astroparticle Physics}\ }\textbf {\bibinfo {volume} {2013}},\ \bibinfo
  {pages} {067} (\bibinfo {year} {2013})}\BibitemShut {NoStop}%
\bibitem [{\citenamefont {Avignone}\ \emph {et~al.}(1998)\citenamefont
  {Avignone}, \citenamefont {Abriola}, \citenamefont {Brodzinski},
  \citenamefont {Collar}, \citenamefont {Creswick}, \citenamefont {DiGregorio},
  \citenamefont {Farach}, \citenamefont {Gattone}, \citenamefont {Gu\'erard},
  \citenamefont {Hasenbalg}, \citenamefont {Huck}, \citenamefont {Miley},
  \citenamefont {Morales}, \citenamefont {Morales}, \citenamefont {Nussinov},
  \citenamefont {Ortiz~de Sol\'orzano}, \citenamefont {Reeves}, \citenamefont
  {Villar},\ and\ \citenamefont {Zioutas}}]{PhysRevLett.81.5068}%
  \BibitemOpen
  \bibfield  {author} {\bibinfo {author} {\bibfnamefont {F.~T.}\ \bibnamefont
  {Avignone}}, \bibinfo {author} {\bibfnamefont {D.}~\bibnamefont {Abriola}},
  \bibinfo {author} {\bibfnamefont {R.~L.}\ \bibnamefont {Brodzinski}},
  \bibinfo {author} {\bibfnamefont {J.~I.}\ \bibnamefont {Collar}}, \bibinfo
  {author} {\bibfnamefont {R.~J.}\ \bibnamefont {Creswick}}, \bibinfo {author}
  {\bibfnamefont {D.~E.}\ \bibnamefont {DiGregorio}}, \bibinfo {author}
  {\bibfnamefont {H.~A.}\ \bibnamefont {Farach}}, \bibinfo {author}
  {\bibfnamefont {A.~O.}\ \bibnamefont {Gattone}}, \bibinfo {author}
  {\bibfnamefont {C.~K.}\ \bibnamefont {Gu\'erard}}, \bibinfo {author}
  {\bibfnamefont {F.}~\bibnamefont {Hasenbalg}}, \bibinfo {author}
  {\bibfnamefont {H.}~\bibnamefont {Huck}}, \bibinfo {author} {\bibfnamefont
  {H.~S.}\ \bibnamefont {Miley}}, \bibinfo {author} {\bibfnamefont
  {A.}~\bibnamefont {Morales}}, \bibinfo {author} {\bibfnamefont
  {J.}~\bibnamefont {Morales}}, \bibinfo {author} {\bibfnamefont
  {S.}~\bibnamefont {Nussinov}}, \bibinfo {author} {\bibfnamefont
  {A.}~\bibnamefont {Ortiz~de Sol\'orzano}}, \bibinfo {author} {\bibfnamefont
  {J.~H.}\ \bibnamefont {Reeves}}, \bibinfo {author} {\bibfnamefont {J.~A.}\
  \bibnamefont {Villar}}, \ and\ \bibinfo {author} {\bibfnamefont
  {K.}~\bibnamefont {Zioutas}} (\bibinfo {collaboration} {SOLAX
  Collaboration}),\ }\href {\doibase 10.1103/PhysRevLett.81.5068} {\bibfield
  {journal} {\bibinfo  {journal} {Phys. Rev. Lett.}\ }\textbf {\bibinfo
  {volume} {81}},\ \bibinfo {pages} {5068} (\bibinfo {year}
  {1998})}\BibitemShut {NoStop}%
\bibitem [{\citenamefont {Morales}\ \emph {et~al.}(2002)\citenamefont {Morales}
  \emph {et~al.}}]{COSME:2001jci}%
  \BibitemOpen
  \bibfield  {author} {\bibinfo {author} {\bibfnamefont {A.}~\bibnamefont
  {Morales}} \emph {et~al.} (\bibinfo {collaboration} {COSME}),\ }\href
  {\doibase 10.1016/S0927-6505(01)00117-7} {\bibfield  {journal} {\bibinfo
  {journal} {Astropart. Phys.}\ }\textbf {\bibinfo {volume} {16}},\ \bibinfo
  {pages} {325} (\bibinfo {year} {2002})},\ \Eprint
  {http://arxiv.org/abs/hep-ex/0101037} {arXiv:hep-ex/0101037} \BibitemShut
  {NoStop}%
\bibitem [{\citenamefont {Ahmed}\ \emph {et~al.}(2009)\citenamefont {Ahmed}
  \emph {et~al.}}]{CDMS:2009fba}%
  \BibitemOpen
  \bibfield  {author} {\bibinfo {author} {\bibfnamefont {Z.}~\bibnamefont
  {Ahmed}} \emph {et~al.} (\bibinfo {collaboration} {CDMS}),\ }\href {\doibase
  10.1103/PhysRevLett.103.141802} {\bibfield  {journal} {\bibinfo  {journal}
  {Phys. Rev. Lett.}\ }\textbf {\bibinfo {volume} {103}},\ \bibinfo {pages}
  {141802} (\bibinfo {year} {2009})},\ \Eprint {http://arxiv.org/abs/0902.4693}
  {arXiv:0902.4693 [hep-ex]} \BibitemShut {NoStop}%
\bibitem [{\citenamefont {Arnquist}\ \emph {et~al.}(2022)\citenamefont
  {Arnquist} \emph {et~al.}}]{Majorana:2022bse}%
  \BibitemOpen
  \bibfield  {author} {\bibinfo {author} {\bibfnamefont {I.~J.}\ \bibnamefont
  {Arnquist}} \emph {et~al.} (\bibinfo {collaboration} {Majorana}),\ }\href
  {\doibase 10.1103/PhysRevLett.129.081803} {\bibfield  {journal} {\bibinfo
  {journal} {Phys. Rev. Lett.}\ }\textbf {\bibinfo {volume} {129}},\ \bibinfo
  {pages} {081803} (\bibinfo {year} {2022})},\ \Eprint
  {http://arxiv.org/abs/2206.05789} {arXiv:2206.05789 [nucl-ex]} \BibitemShut
  {NoStop}%
\bibitem [{\citenamefont {Albakry}\ \emph {et~al.}(2022)\citenamefont {Albakry}
  \emph {et~al.}}]{SuperCDMS:2022kse}%
  \BibitemOpen
  \bibfield  {author} {\bibinfo {author} {\bibfnamefont {M.~F.}\ \bibnamefont
  {Albakry}} \emph {et~al.} (\bibinfo {collaboration} {SuperCDMS}),\ }in\
  \href@noop {} {\emph {\bibinfo {booktitle} {{2022 Snowmass Summer Study}}}}\
  (\bibinfo {year} {2022})\ \Eprint {http://arxiv.org/abs/2203.08463}
  {arXiv:2203.08463 [physics.ins-det]} \BibitemShut {NoStop}%
\bibitem [{\citenamefont {Abgrall}\ \emph {et~al.}(2021)\citenamefont {Abgrall}
  \emph {et~al.}}]{LEGEND:2021bnm}%
  \BibitemOpen
  \bibfield  {author} {\bibinfo {author} {\bibfnamefont {N.}~\bibnamefont
  {Abgrall}} \emph {et~al.} (\bibinfo {collaboration} {LEGEND}),\ }\href@noop
  {} {\  (\bibinfo {year} {2021})},\ \Eprint {http://arxiv.org/abs/2107.11462}
  {arXiv:2107.11462 [physics.ins-det]} \BibitemShut {NoStop}%
\bibitem [{\citenamefont {Antonello}\ \emph {et~al.}(2019)\citenamefont
  {Antonello} \emph {et~al.}}]{SABRE:2018lfp}%
  \BibitemOpen
  \bibfield  {author} {\bibinfo {author} {\bibfnamefont {M.}~\bibnamefont
  {Antonello}} \emph {et~al.} (\bibinfo {collaboration} {SABRE}),\ }\href
  {\doibase 10.1140/epjc/s10052-019-6860-y} {\bibfield  {journal} {\bibinfo
  {journal} {Eur. Phys. J. C}\ }\textbf {\bibinfo {volume} {79}},\ \bibinfo
  {pages} {363} (\bibinfo {year} {2019})},\ \Eprint
  {http://arxiv.org/abs/1806.09340} {arXiv:1806.09340 [physics.ins-det]}
  \BibitemShut {NoStop}%
\bibitem [{\citenamefont {Buchm{\"u}ller}\ and\ \citenamefont
  {Hoogeveen}(1990)}]{Buchmuller:1989rb}%
  \BibitemOpen
  \bibfield  {author} {\bibinfo {author} {\bibfnamefont {W.}~\bibnamefont
  {Buchm{\"u}ller}}\ and\ \bibinfo {author} {\bibfnamefont {F.}~\bibnamefont
  {Hoogeveen}},\ }\href {\doibase 10.1016/0370-2693(90)91444-G} {\bibfield
  {journal} {\bibinfo  {journal} {Phys.\ Lett.\ B}\ }\textbf {\bibinfo {volume}
  {237}},\ \bibinfo {pages} {278} (\bibinfo {year} {1990})}\BibitemShut
  {NoStop}%
\bibitem [{\citenamefont {Warren}(1969)}]{warren}%
  \BibitemOpen
  \bibfield  {author} {\bibinfo {author} {\bibfnamefont {B.~E.}\ \bibnamefont
  {Warren}},\ }\href@noop {} {\emph {\bibinfo {title} {X-Ray diffraction}}}\
  (\bibinfo  {publisher} {Addison Wesley},\ \bibinfo {year} {1969})\BibitemShut
  {NoStop}%
\bibitem [{\citenamefont {Yamaji}\ \emph {et~al.}(2017)\citenamefont {Yamaji},
  \citenamefont {Yamazaki}, \citenamefont {Tamasaku},\ and\ \citenamefont
  {Namba}}]{Yamaji:2017pep}%
  \BibitemOpen
  \bibfield  {author} {\bibinfo {author} {\bibfnamefont {T.}~\bibnamefont
  {Yamaji}}, \bibinfo {author} {\bibfnamefont {T.}~\bibnamefont {Yamazaki}},
  \bibinfo {author} {\bibfnamefont {K.}~\bibnamefont {Tamasaku}}, \ and\
  \bibinfo {author} {\bibfnamefont {T.}~\bibnamefont {Namba}},\ }\href
  {\doibase 10.1103/PhysRevD.96.115001} {\bibfield  {journal} {\bibinfo
  {journal} {Phys. Rev. D}\ }\textbf {\bibinfo {volume} {96}},\ \bibinfo
  {pages} {115001} (\bibinfo {year} {2017})},\ \Eprint
  {http://arxiv.org/abs/1709.03299} {arXiv:1709.03299 [physics.ins-det]}
  \BibitemShut {NoStop}%
\bibitem [{\citenamefont {Borrmann}(1954)}]{borrmann1954}%
  \BibitemOpen
  \bibfield  {author} {\bibinfo {author} {\bibfnamefont {G.}~\bibnamefont
  {Borrmann}},\ }\href {\doibase doi:10.1524/zkri.1954.106.16.109} {\bibfield
  {journal} {\bibinfo  {journal} {Zeitschrift f\"{u}r Kristallographie -
  Crystalline Materials}\ }\textbf {\bibinfo {volume} {106}},\ \bibinfo {pages}
  {109} (\bibinfo {year} {1954})}\BibitemShut {NoStop}%
\bibitem [{\citenamefont {Zachariasen}(1945)}]{zachariasenBook}%
  \BibitemOpen
  \bibfield  {author} {\bibinfo {author} {\bibfnamefont {W.~H.}\ \bibnamefont
  {Zachariasen}},\ }\href@noop {} {\emph {\bibinfo {title} {The Theory of X-ray
  Diffraction in Crystals}}}\ (\bibinfo  {publisher} {John Wiley \& Sons, Inc.,
  New York},\ \bibinfo {year} {1945})\BibitemShut {NoStop}%
\bibitem [{\citenamefont {Zachariasen}(1952)}]{zachariasenPNAS}%
  \BibitemOpen
  \bibfield  {author} {\bibinfo {author} {\bibfnamefont {W.~H.}\ \bibnamefont
  {Zachariasen}},\ }\href {\doibase 10.1073/pnas.38.4.378} {\bibfield
  {journal} {\bibinfo  {journal} {Proceedings of the National Academy of
  Sciences}\ }\textbf {\bibinfo {volume} {38}},\ \bibinfo {pages} {378}
  (\bibinfo {year} {1952})},\ \Eprint
  {http://arxiv.org/abs/https://www.pnas.org/doi/pdf/10.1073/pnas.38.4.378}
  {https://www.pnas.org/doi/pdf/10.1073/pnas.38.4.378} \BibitemShut {NoStop}%
\bibitem [{\citenamefont {Batterman}(1961)}]{batterman1}%
  \BibitemOpen
  \bibfield  {author} {\bibinfo {author} {\bibfnamefont {B.~W.}\ \bibnamefont
  {Batterman}},\ }\href {\doibase 10.1063/1.1736201} {\bibfield  {journal}
  {\bibinfo  {journal} {Journal of Applied Physics}\ }\textbf {\bibinfo
  {volume} {32}},\ \bibinfo {pages} {998} (\bibinfo {year} {1961})},\ \Eprint
  {http://arxiv.org/abs/https://doi.org/10.1063/1.1736201}
  {https://doi.org/10.1063/1.1736201} \BibitemShut {NoStop}%
\bibitem [{\citenamefont {Batterman}(1962)}]{batterman2}%
  \BibitemOpen
  \bibfield  {author} {\bibinfo {author} {\bibfnamefont {B.~W.}\ \bibnamefont
  {Batterman}},\ }\href {\doibase 10.1103/PhysRev.126.1461} {\bibfield
  {journal} {\bibinfo  {journal} {Phys. Rev.}\ }\textbf {\bibinfo {volume}
  {126}},\ \bibinfo {pages} {1461} (\bibinfo {year} {1962})}\BibitemShut
  {NoStop}%
\bibitem [{\citenamefont {Hirsch}(1952)}]{Hirsch:a00588}%
  \BibitemOpen
  \bibfield  {author} {\bibinfo {author} {\bibfnamefont {P.~B.}\ \bibnamefont
  {Hirsch}},\ }\href {\doibase 10.1107/S0365110X52000526} {\bibfield  {journal}
  {\bibinfo  {journal} {Acta Crystallographica}\ }\textbf {\bibinfo {volume}
  {5}},\ \bibinfo {pages} {176} (\bibinfo {year} {1952})}\BibitemShut {NoStop}%
\bibitem [{\citenamefont {Biagini}(1990)}]{PhysRevA.42.3695}%
  \BibitemOpen
  \bibfield  {author} {\bibinfo {author} {\bibfnamefont {M.}~\bibnamefont
  {Biagini}},\ }\href {\doibase 10.1103/PhysRevA.42.3695} {\bibfield  {journal}
  {\bibinfo  {journal} {Phys. Rev. A}\ }\textbf {\bibinfo {volume} {42}},\
  \bibinfo {pages} {3695} (\bibinfo {year} {1990})}\BibitemShut {NoStop}%
\bibitem [{\citenamefont {Biagini}(1991)}]{PhysRevA.44.645}%
  \BibitemOpen
  \bibfield  {author} {\bibinfo {author} {\bibfnamefont {M.}~\bibnamefont
  {Biagini}},\ }\href {\doibase 10.1103/PhysRevA.44.645} {\bibfield  {journal}
  {\bibinfo  {journal} {Phys. Rev. A}\ }\textbf {\bibinfo {volume} {44}},\
  \bibinfo {pages} {645} (\bibinfo {year} {1991})}\BibitemShut {NoStop}%
\bibitem [{\citenamefont {Poshakinskiy}\ and\ \citenamefont
  {Poddubny}(2021)}]{Poshakinskiy_2021}%
  \BibitemOpen
  \bibfield  {author} {\bibinfo {author} {\bibfnamefont {A.~V.}\ \bibnamefont
  {Poshakinskiy}}\ and\ \bibinfo {author} {\bibfnamefont {A.~N.}\ \bibnamefont
  {Poddubny}},\ }\href {\doibase 10.1103/physreva.103.043718} {\bibfield
  {journal} {\bibinfo  {journal} {Physical Review A}\ }\textbf {\bibinfo
  {volume} {103}} (\bibinfo {year} {2021}),\
  10.1103/physreva.103.043718}\BibitemShut {NoStop}%
\bibitem [{\citenamefont {Pettifer}\ \emph {et~al.}(2008)\citenamefont
  {Pettifer}, \citenamefont {Collins},\ and\ \citenamefont
  {Laundy}}]{pettifer_collins_laundy_2008}%
  \BibitemOpen
  \bibfield  {author} {\bibinfo {author} {\bibfnamefont {R.~F.}\ \bibnamefont
  {Pettifer}}, \bibinfo {author} {\bibfnamefont {S.~P.}\ \bibnamefont
  {Collins}}, \ and\ \bibinfo {author} {\bibfnamefont {D.}~\bibnamefont
  {Laundy}},\ }\href {\doibase 10.1038/nature07099} {\bibfield  {journal}
  {\bibinfo  {journal} {Nature}\ }\textbf {\bibinfo {volume} {454}},\ \bibinfo
  {pages} {196–199} (\bibinfo {year} {2008})}\BibitemShut {NoStop}%
\bibitem [{\citenamefont {Cebrian}\ \emph {et~al.}(1999)\citenamefont {Cebrian}
  \emph {et~al.}}]{Cebrian:1998mu}%
  \BibitemOpen
  \bibfield  {author} {\bibinfo {author} {\bibfnamefont {S.}~\bibnamefont
  {Cebrian}} \emph {et~al.},\ }\href {\doibase 10.1016/S0927-6505(98)00069-3}
  {\bibfield  {journal} {\bibinfo  {journal} {Astropart. Phys.}\ }\textbf
  {\bibinfo {volume} {10}},\ \bibinfo {pages} {397} (\bibinfo {year} {1999})},\
  \Eprint {http://arxiv.org/abs/astro-ph/9811359} {arXiv:astro-ph/9811359}
  \BibitemShut {NoStop}%
\bibitem [{\citenamefont {Bernabei}\ \emph {et~al.}(2001)\citenamefont
  {Bernabei} \emph {et~al.}}]{Bernabei:2001ny}%
  \BibitemOpen
  \bibfield  {author} {\bibinfo {author} {\bibfnamefont {R.}~\bibnamefont
  {Bernabei}} \emph {et~al.},\ }\href {\doibase 10.1016/S0370-2693(01)00840-1}
  {\bibfield  {journal} {\bibinfo  {journal} {Phys. Lett. B}\ }\textbf
  {\bibinfo {volume} {515}},\ \bibinfo {pages} {6} (\bibinfo {year}
  {2001})}\BibitemShut {NoStop}%
\bibitem [{\citenamefont {Bednyakov}\ and\ \citenamefont
  {Naumov}(2021)}]{Bednyakov:2021lul}%
  \BibitemOpen
  \bibfield  {author} {\bibinfo {author} {\bibfnamefont {V.~A.}\ \bibnamefont
  {Bednyakov}}\ and\ \bibinfo {author} {\bibfnamefont {D.~V.}\ \bibnamefont
  {Naumov}},\ }\href {\doibase 10.1134/S1063779620060039} {\bibfield  {journal}
  {\bibinfo  {journal} {Phys. Part. Nucl.}\ }\textbf {\bibinfo {volume} {52}},\
  \bibinfo {pages} {39} (\bibinfo {year} {2021})}\BibitemShut {NoStop}%
\bibitem [{\citenamefont {Tsai}(1986)}]{Tsai:1986tx}%
  \BibitemOpen
  \bibfield  {author} {\bibinfo {author} {\bibfnamefont {Y.-S.}\ \bibnamefont
  {Tsai}},\ }\bibfield  {booktitle} {\emph {\bibinfo {booktitle} {{Proceedings,
  23RD International Conference on High Energy Physics, JULY 16-23, 1986,
  Berkeley, CA}}},\ }\href {\doibase 10.1103/PhysRevD.34.1326} {\bibfield
  {journal} {\bibinfo  {journal} {Phys. Rev.}\ }\textbf {\bibinfo {volume}
  {D34}},\ \bibinfo {pages} {1326} (\bibinfo {year} {1986})}\BibitemShut
  {NoStop}%
\bibitem [{\citenamefont {Tsai}(1974)}]{RevModPhys.46.815}%
  \BibitemOpen
  \bibfield  {author} {\bibinfo {author} {\bibfnamefont {Y.-S.}\ \bibnamefont
  {Tsai}},\ }\href {\doibase 10.1103/RevModPhys.46.815} {\bibfield  {journal}
  {\bibinfo  {journal} {Rev. Mod. Phys.}\ }\textbf {\bibinfo {volume} {46}},\
  \bibinfo {pages} {815} (\bibinfo {year} {1974})}\BibitemShut {NoStop}%
\bibitem [{\citenamefont {Chantler}(2000)}]{chantler_NIST}%
  \BibitemOpen
  \bibfield  {author} {\bibinfo {author} {\bibfnamefont {C.~T.}\ \bibnamefont
  {Chantler}},\ }\href {\doibase 10.1063/1.1321055} {\bibfield  {journal}
  {\bibinfo  {journal} {Journal of Physical and Chemical Reference Data}\
  }\textbf {\bibinfo {volume} {29}},\ \bibinfo {pages} {597} (\bibinfo {year}
  {2000})},\ \Eprint {http://arxiv.org/abs/https://doi.org/10.1063/1.1321055}
  {https://doi.org/10.1063/1.1321055} \BibitemShut {NoStop}%
\bibitem [{\citenamefont {Wagenfeld}(1987)}]{wagenfield1986}%
  \BibitemOpen
  \bibfield  {author} {\bibinfo {author} {\bibfnamefont {H.~K.}\ \bibnamefont
  {Wagenfeld}},\ }\href {\doibase 10.1007/BF01303765} {\bibfield  {journal}
  {\bibinfo  {journal} {Zeitschrift f{\"u}r Physik B Condensed Matter}\
  }\textbf {\bibinfo {volume} {65}},\ \bibinfo {pages} {437} (\bibinfo {year}
  {1987})}\BibitemShut {NoStop}%
\bibitem [{\citenamefont {Wagenfeld}(1966)}]{PhysRev.144.216}%
  \BibitemOpen
  \bibfield  {author} {\bibinfo {author} {\bibfnamefont {H.}~\bibnamefont
  {Wagenfeld}},\ }\href {\doibase 10.1103/PhysRev.144.216} {\bibfield
  {journal} {\bibinfo  {journal} {Phys. Rev.}\ }\textbf {\bibinfo {volume}
  {144}},\ \bibinfo {pages} {216} (\bibinfo {year} {1966})}\BibitemShut
  {NoStop}%
\bibitem [{\citenamefont {Persson}\ and\ \citenamefont
  {Efimov}(1970)}]{persson_efimov}%
  \BibitemOpen
  \bibfield  {author} {\bibinfo {author} {\bibfnamefont {E.}~\bibnamefont
  {Persson}}\ and\ \bibinfo {author} {\bibfnamefont {O.~N.}\ \bibnamefont
  {Efimov}},\ }\href {\doibase https://doi.org/10.1002/pssa.19700020412}
  {\bibfield  {journal} {\bibinfo  {journal} {physica status solidi (a)}\
  }\textbf {\bibinfo {volume} {2}},\ \bibinfo {pages} {757} (\bibinfo {year}
  {1970})}\BibitemShut {NoStop}%
\bibitem [{\citenamefont {Peng}\ \emph {et~al.}(1996)\citenamefont {Peng},
  \citenamefont {Ren}, \citenamefont {Dudarev},\ and\ \citenamefont
  {Whelan}}]{Peng96}%
  \BibitemOpen
  \bibfield  {author} {\bibinfo {author} {\bibfnamefont {L.-M.}\ \bibnamefont
  {Peng}}, \bibinfo {author} {\bibfnamefont {G.}~\bibnamefont {Ren}}, \bibinfo
  {author} {\bibfnamefont {S.~L.}\ \bibnamefont {Dudarev}}, \ and\ \bibinfo
  {author} {\bibfnamefont {M.~J.}\ \bibnamefont {Whelan}},\ }\href {\doibase
  https://doi.org/10.1107/S010876739600089X} {\bibfield  {journal} {\bibinfo
  {journal} {Acta Crystallographica Section A}\ }\textbf {\bibinfo {volume}
  {52}},\ \bibinfo {pages} {456} (\bibinfo {year} {1996})}\BibitemShut
  {NoStop}%
\bibitem [{\citenamefont {Catena}\ \emph {et~al.}(2020)\citenamefont {Catena},
  \citenamefont {Emken}, \citenamefont {Spaldin},\ and\ \citenamefont
  {Tarantino}}]{Catena:2019gfa}%
  \BibitemOpen
  \bibfield  {author} {\bibinfo {author} {\bibfnamefont {R.}~\bibnamefont
  {Catena}}, \bibinfo {author} {\bibfnamefont {T.}~\bibnamefont {Emken}},
  \bibinfo {author} {\bibfnamefont {N.~A.}\ \bibnamefont {Spaldin}}, \ and\
  \bibinfo {author} {\bibfnamefont {W.}~\bibnamefont {Tarantino}},\ }\href
  {\doibase 10.1103/PhysRevResearch.2.033195} {\bibfield  {journal} {\bibinfo
  {journal} {Phys. Rev. Res.}\ }\textbf {\bibinfo {volume} {2}},\ \bibinfo
  {pages} {033195} (\bibinfo {year} {2020})},\ \Eprint
  {http://arxiv.org/abs/1912.08204} {arXiv:1912.08204 [hep-ph]} \BibitemShut
  {NoStop}%
\bibitem [{\citenamefont {Redondo}(2013)}]{Redondo:2013wwa}%
  \BibitemOpen
  \bibfield  {author} {\bibinfo {author} {\bibfnamefont {J.}~\bibnamefont
  {Redondo}},\ }\href {\doibase 10.1088/1475-7516/2013/12/008} {\bibfield
  {journal} {\bibinfo  {journal} {JCAP}\ }\textbf {\bibinfo {volume} {12}},\
  \bibinfo {pages} {008} (\bibinfo {year} {2013})},\ \Eprint
  {http://arxiv.org/abs/1310.0823} {arXiv:1310.0823 [hep-ph]} \BibitemShut
  {NoStop}%
\bibitem [{\citenamefont {Dicus}\ \emph {et~al.}(1978)\citenamefont {Dicus},
  \citenamefont {Kolb}, \citenamefont {Teplitz},\ and\ \citenamefont
  {Wagoner}}]{PhysRevD.18.1829}%
  \BibitemOpen
  \bibfield  {author} {\bibinfo {author} {\bibfnamefont {D.~A.}\ \bibnamefont
  {Dicus}}, \bibinfo {author} {\bibfnamefont {E.~W.}\ \bibnamefont {Kolb}},
  \bibinfo {author} {\bibfnamefont {V.~L.}\ \bibnamefont {Teplitz}}, \ and\
  \bibinfo {author} {\bibfnamefont {R.~V.}\ \bibnamefont {Wagoner}},\ }\href
  {\doibase 10.1103/PhysRevD.18.1829} {\bibfield  {journal} {\bibinfo
  {journal} {Phys. Rev. D}\ }\textbf {\bibinfo {volume} {18}},\ \bibinfo
  {pages} {1829} (\bibinfo {year} {1978})}\BibitemShut {NoStop}%
\bibitem [{\citenamefont {Di~Lella}\ \emph {et~al.}(2000)\citenamefont
  {Di~Lella}, \citenamefont {Pilaftsis}, \citenamefont {Raffelt},\ and\
  \citenamefont {Zioutas}}]{DiLella:2000dn}%
  \BibitemOpen
  \bibfield  {author} {\bibinfo {author} {\bibfnamefont {L.}~\bibnamefont
  {Di~Lella}}, \bibinfo {author} {\bibfnamefont {A.}~\bibnamefont {Pilaftsis}},
  \bibinfo {author} {\bibfnamefont {G.}~\bibnamefont {Raffelt}}, \ and\
  \bibinfo {author} {\bibfnamefont {K.}~\bibnamefont {Zioutas}},\ }\href
  {\doibase 10.1103/PhysRevD.62.125011} {\bibfield  {journal} {\bibinfo
  {journal} {Phys. Rev. D}\ }\textbf {\bibinfo {volume} {62}},\ \bibinfo
  {pages} {125011} (\bibinfo {year} {2000})},\ \Eprint
  {http://arxiv.org/abs/hep-ph/0006327} {arXiv:hep-ph/0006327} \BibitemShut
  {NoStop}%
\bibitem [{\citenamefont {Andriamonje}\ \emph {et~al.}(2007)\citenamefont
  {Andriamonje} \emph {et~al.}}]{CAST:2007jps}%
  \BibitemOpen
  \bibfield  {author} {\bibinfo {author} {\bibfnamefont {S.}~\bibnamefont
  {Andriamonje}} \emph {et~al.} (\bibinfo {collaboration} {CAST}),\ }\href
  {\doibase 10.1088/1475-7516/2007/04/010} {\bibfield  {journal} {\bibinfo
  {journal} {JCAP}\ }\textbf {\bibinfo {volume} {04}},\ \bibinfo {pages} {010}
  (\bibinfo {year} {2007})},\ \Eprint {http://arxiv.org/abs/hep-ex/0702006}
  {arXiv:hep-ex/0702006} \BibitemShut {NoStop}%
\bibitem [{\citenamefont {Reda}\ and\ \citenamefont {Andreas}(2008)}]{nrel}%
  \BibitemOpen
  \bibfield  {author} {\bibinfo {author} {\bibfnamefont {I.}~\bibnamefont
  {Reda}}\ and\ \bibinfo {author} {\bibfnamefont {A.}~\bibnamefont {Andreas}},\
  }\href@noop {} {\emph {\bibinfo {title} {{Solar Position Algorithm for Solar
  Radiation Applications}}}},\ \bibinfo {type} {Tech. Rep.}\ (\bibinfo
  {institution} {U.S. Department of Energy},\ \bibinfo {year}
  {2008})\BibitemShut {NoStop}%
\bibitem [{\citenamefont {Kim}(1979)}]{PhysRevLett.43.103}%
  \BibitemOpen
  \bibfield  {author} {\bibinfo {author} {\bibfnamefont {J.~E.}\ \bibnamefont
  {Kim}},\ }\href {\doibase 10.1103/PhysRevLett.43.103} {\bibfield  {journal}
  {\bibinfo  {journal} {Phys. Rev. Lett.}\ }\textbf {\bibinfo {volume} {43}},\
  \bibinfo {pages} {103} (\bibinfo {year} {1979})}\BibitemShut {NoStop}%
\bibitem [{\citenamefont {Shifman}\ \emph {et~al.}(1980)\citenamefont
  {Shifman}, \citenamefont {Vainshtein},\ and\ \citenamefont
  {Zakharov}}]{SHIFMAN1980493}%
  \BibitemOpen
  \bibfield  {author} {\bibinfo {author} {\bibfnamefont {M.}~\bibnamefont
  {Shifman}}, \bibinfo {author} {\bibfnamefont {A.}~\bibnamefont {Vainshtein}},
  \ and\ \bibinfo {author} {\bibfnamefont {V.}~\bibnamefont {Zakharov}},\
  }\href {\doibase https://doi.org/10.1016/0550-3213(80)90209-6} {\bibfield
  {journal} {\bibinfo  {journal} {Nuclear Physics B}\ }\textbf {\bibinfo
  {volume} {166}},\ \bibinfo {pages} {493} (\bibinfo {year}
  {1980})}\BibitemShut {NoStop}%
\bibitem [{\citenamefont {Zhitnitsky}(1980)}]{Zhitnitsky:1980tq}%
  \BibitemOpen
  \bibfield  {author} {\bibinfo {author} {\bibfnamefont {A.~R.}\ \bibnamefont
  {Zhitnitsky}},\ }\href@noop {} {\bibfield  {journal} {\bibinfo  {journal}
  {Sov. J. Nucl. Phys.}\ }\textbf {\bibinfo {volume} {31}},\ \bibinfo {pages}
  {260} (\bibinfo {year} {1980})}\BibitemShut {NoStop}%
\bibitem [{\citenamefont {Dine}\ \emph
  {et~al.}(1981{\natexlab{a}})\citenamefont {Dine}, \citenamefont {Fischler},\
  and\ \citenamefont {Srednicki}}]{DINE1981199}%
  \BibitemOpen
  \bibfield  {author} {\bibinfo {author} {\bibfnamefont {M.}~\bibnamefont
  {Dine}}, \bibinfo {author} {\bibfnamefont {W.}~\bibnamefont {Fischler}}, \
  and\ \bibinfo {author} {\bibfnamefont {M.}~\bibnamefont {Srednicki}},\ }\href
  {\doibase https://doi.org/10.1016/0370-2693(81)90590-6} {\bibfield  {journal}
  {\bibinfo  {journal} {Physics Letters B}\ }\textbf {\bibinfo {volume}
  {104}},\ \bibinfo {pages} {199} (\bibinfo {year}
  {1981}{\natexlab{a}})}\BibitemShut {NoStop}%
\bibitem [{\citenamefont {Dine}\ \emph
  {et~al.}(1981{\natexlab{b}})\citenamefont {Dine}, \citenamefont {Fischler},\
  and\ \citenamefont {Srednicki}}]{Dine:1981rt}%
  \BibitemOpen
  \bibfield  {author} {\bibinfo {author} {\bibfnamefont {M.}~\bibnamefont
  {Dine}}, \bibinfo {author} {\bibfnamefont {W.}~\bibnamefont {Fischler}}, \
  and\ \bibinfo {author} {\bibfnamefont {M.}~\bibnamefont {Srednicki}},\ }\href
  {\doibase 10.1016/0370-2693(81)90590-6} {\bibfield  {journal} {\bibinfo
  {journal} {Phys. Lett. B}\ }\textbf {\bibinfo {volume} {104}},\ \bibinfo
  {pages} {199} (\bibinfo {year} {1981}{\natexlab{b}})}\BibitemShut {NoStop}%
\bibitem [{\citenamefont {Di~Luzio}\ \emph {et~al.}(2020)\citenamefont
  {Di~Luzio}, \citenamefont {Giannotti}, \citenamefont {Nardi},\ and\
  \citenamefont {Visinelli}}]{DiLuzio:2020wdo}%
  \BibitemOpen
  \bibfield  {author} {\bibinfo {author} {\bibfnamefont {L.}~\bibnamefont
  {Di~Luzio}}, \bibinfo {author} {\bibfnamefont {M.}~\bibnamefont {Giannotti}},
  \bibinfo {author} {\bibfnamefont {E.}~\bibnamefont {Nardi}}, \ and\ \bibinfo
  {author} {\bibfnamefont {L.}~\bibnamefont {Visinelli}},\ }\href {\doibase
  10.1016/j.physrep.2020.06.002} {\bibfield  {journal} {\bibinfo  {journal}
  {Phys. Rept.}\ }\textbf {\bibinfo {volume} {870}},\ \bibinfo {pages} {1}
  (\bibinfo {year} {2020})},\ \Eprint {http://arxiv.org/abs/2003.01100}
  {arXiv:2003.01100 [hep-ph]} \BibitemShut {NoStop}%
\bibitem [{\citenamefont {Kivel}\ \emph {et~al.}(2022)\citenamefont {Kivel},
  \citenamefont {Laux},\ and\ \citenamefont {Yu}}]{Kivel:2022emq}%
  \BibitemOpen
  \bibfield  {author} {\bibinfo {author} {\bibfnamefont {A.}~\bibnamefont
  {Kivel}}, \bibinfo {author} {\bibfnamefont {J.}~\bibnamefont {Laux}}, \ and\
  \bibinfo {author} {\bibfnamefont {F.}~\bibnamefont {Yu}},\ }\href {\doibase
  10.1007/JHEP11(2022)088} {\bibfield  {journal} {\bibinfo  {journal} {JHEP}\
  }\textbf {\bibinfo {volume} {11}},\ \bibinfo {pages} {088} (\bibinfo {year}
  {2022})},\ \Eprint {http://arxiv.org/abs/2207.08740} {arXiv:2207.08740
  [hep-ph]} \BibitemShut {NoStop}%
\bibitem [{\citenamefont {Valenti}\ \emph {et~al.}(2022)\citenamefont
  {Valenti}, \citenamefont {Vecchi},\ and\ \citenamefont
  {Xu}}]{Valenti:2022tsc}%
  \BibitemOpen
  \bibfield  {author} {\bibinfo {author} {\bibfnamefont {A.}~\bibnamefont
  {Valenti}}, \bibinfo {author} {\bibfnamefont {L.}~\bibnamefont {Vecchi}}, \
  and\ \bibinfo {author} {\bibfnamefont {L.-X.}\ \bibnamefont {Xu}},\ }\href
  {\doibase 10.1007/JHEP10(2022)025} {\bibfield  {journal} {\bibinfo  {journal}
  {JHEP}\ }\textbf {\bibinfo {volume} {10}},\ \bibinfo {pages} {025} (\bibinfo
  {year} {2022})},\ \Eprint {http://arxiv.org/abs/2206.04077} {arXiv:2206.04077
  [hep-ph]} \BibitemShut {NoStop}%
\bibitem [{\citenamefont {Giannotti}\ \emph {et~al.}(2016)\citenamefont
  {Giannotti}, \citenamefont {Irastorza}, \citenamefont {Redondo},\ and\
  \citenamefont {Ringwald}}]{Giannotti:2015kwo}%
  \BibitemOpen
  \bibfield  {author} {\bibinfo {author} {\bibfnamefont {M.}~\bibnamefont
  {Giannotti}}, \bibinfo {author} {\bibfnamefont {I.}~\bibnamefont
  {Irastorza}}, \bibinfo {author} {\bibfnamefont {J.}~\bibnamefont {Redondo}},
  \ and\ \bibinfo {author} {\bibfnamefont {A.}~\bibnamefont {Ringwald}},\
  }\href {\doibase 10.1088/1475-7516/2016/05/057} {\bibfield  {journal}
  {\bibinfo  {journal} {JCAP}\ }\textbf {\bibinfo {volume} {05}},\ \bibinfo
  {pages} {057} (\bibinfo {year} {2016})},\ \Eprint
  {http://arxiv.org/abs/1512.08108} {arXiv:1512.08108 [astro-ph.HE]}
  \BibitemShut {NoStop}%
\bibitem [{\citenamefont {Hoof}\ \emph {et~al.}(2019)\citenamefont {Hoof},
  \citenamefont {Kahlhoefer}, \citenamefont {Scott}, \citenamefont {Weniger},\
  and\ \citenamefont {White}}]{Hoof:2018ieb}%
  \BibitemOpen
  \bibfield  {author} {\bibinfo {author} {\bibfnamefont {S.}~\bibnamefont
  {Hoof}}, \bibinfo {author} {\bibfnamefont {F.}~\bibnamefont {Kahlhoefer}},
  \bibinfo {author} {\bibfnamefont {P.}~\bibnamefont {Scott}}, \bibinfo
  {author} {\bibfnamefont {C.}~\bibnamefont {Weniger}}, \ and\ \bibinfo
  {author} {\bibfnamefont {M.}~\bibnamefont {White}},\ }\href {\doibase
  10.1007/JHEP03(2019)191} {\bibfield  {journal} {\bibinfo  {journal} {JHEP}\
  }\textbf {\bibinfo {volume} {03}},\ \bibinfo {pages} {191} (\bibinfo {year}
  {2019})},\ \bibinfo {note} {[Erratum: JHEP 11, 099 (2019)]},\ \Eprint
  {http://arxiv.org/abs/1810.07192} {arXiv:1810.07192 [hep-ph]} \BibitemShut
  {NoStop}%
\bibitem [{\citenamefont {Ayala}\ \emph {et~al.}(2014)\citenamefont {Ayala},
  \citenamefont {Dom\'\i{}nguez}, \citenamefont {Giannotti}, \citenamefont
  {Mirizzi},\ and\ \citenamefont {Straniero}}]{Ayala:2014pea}%
  \BibitemOpen
  \bibfield  {author} {\bibinfo {author} {\bibfnamefont {A.}~\bibnamefont
  {Ayala}}, \bibinfo {author} {\bibfnamefont {I.}~\bibnamefont
  {Dom\'\i{}nguez}}, \bibinfo {author} {\bibfnamefont {M.}~\bibnamefont
  {Giannotti}}, \bibinfo {author} {\bibfnamefont {A.}~\bibnamefont {Mirizzi}},
  \ and\ \bibinfo {author} {\bibfnamefont {O.}~\bibnamefont {Straniero}},\
  }\href {\doibase 10.1103/PhysRevLett.113.191302} {\bibfield  {journal}
  {\bibinfo  {journal} {Phys. Rev. Lett.}\ }\textbf {\bibinfo {volume} {113}},\
  \bibinfo {pages} {191302} (\bibinfo {year} {2014})},\ \Eprint
  {http://arxiv.org/abs/1406.6053} {arXiv:1406.6053 [astro-ph.SR]} \BibitemShut
  {NoStop}%
\bibitem [{\citenamefont {Dent}\ \emph {et~al.}(2020)\citenamefont {Dent},
  \citenamefont {Dutta}, \citenamefont {Newstead},\ and\ \citenamefont
  {Thompson}}]{Dent:2020jhf}%
  \BibitemOpen
  \bibfield  {author} {\bibinfo {author} {\bibfnamefont {J.~B.}\ \bibnamefont
  {Dent}}, \bibinfo {author} {\bibfnamefont {B.}~\bibnamefont {Dutta}},
  \bibinfo {author} {\bibfnamefont {J.~L.}\ \bibnamefont {Newstead}}, \ and\
  \bibinfo {author} {\bibfnamefont {A.}~\bibnamefont {Thompson}},\ }\href
  {\doibase 10.1103/PhysRevLett.125.131805} {\bibfield  {journal} {\bibinfo
  {journal} {Phys. Rev. Lett.}\ }\textbf {\bibinfo {volume} {125}},\ \bibinfo
  {pages} {131805} (\bibinfo {year} {2020})},\ \Eprint
  {http://arxiv.org/abs/2006.15118} {arXiv:2006.15118 [hep-ph]} \BibitemShut
  {NoStop}%
\bibitem [{\citenamefont {Hardy}\ and\ \citenamefont
  {Lasenby}(2017)}]{Hardy:2016kme}%
  \BibitemOpen
  \bibfield  {author} {\bibinfo {author} {\bibfnamefont {E.}~\bibnamefont
  {Hardy}}\ and\ \bibinfo {author} {\bibfnamefont {R.}~\bibnamefont
  {Lasenby}},\ }\href {\doibase 10.1007/JHEP02(2017)033} {\bibfield  {journal}
  {\bibinfo  {journal} {JHEP}\ }\textbf {\bibinfo {volume} {02}},\ \bibinfo
  {pages} {033} (\bibinfo {year} {2017})},\ \Eprint
  {http://arxiv.org/abs/1611.05852} {arXiv:1611.05852 [hep-ph]} \BibitemShut
  {NoStop}%
\bibitem [{\citenamefont {Aprile}\ \emph {et~al.}(2022)\citenamefont {Aprile}
  \emph {et~al.}}]{XENON:2022mpc}%
  \BibitemOpen
  \bibfield  {author} {\bibinfo {author} {\bibfnamefont {E.}~\bibnamefont
  {Aprile}} \emph {et~al.} (\bibinfo {collaboration} {XENON}),\ }\href@noop {}
  {\  (\bibinfo {year} {2022})},\ \Eprint {http://arxiv.org/abs/2207.11330}
  {arXiv:2207.11330 [hep-ex]} \BibitemShut {NoStop}%
\bibitem [{\citenamefont {Armengaud}\ \emph {et~al.}(2019)\citenamefont
  {Armengaud} \emph {et~al.}}]{Armengaud:2019uso}%
  \BibitemOpen
  \bibfield  {author} {\bibinfo {author} {\bibfnamefont {E.}~\bibnamefont
  {Armengaud}} \emph {et~al.} (\bibinfo {collaboration} {IAXO}),\ }\href
  {\doibase 10.1088/1475-7516/2019/06/047} {\bibfield  {journal} {\bibinfo
  {journal} {JCAP}\ }\textbf {\bibinfo {volume} {06}},\ \bibinfo {pages} {047}
  (\bibinfo {year} {2019})},\ \Eprint {http://arxiv.org/abs/1904.09155}
  {arXiv:1904.09155 [hep-ph]} \BibitemShut {NoStop}%
\bibitem [{\citenamefont {Benato}\ \emph {et~al.}(2019)\citenamefont {Benato},
  \citenamefont {Drobizhev}, \citenamefont {Rajendran},\ and\ \citenamefont
  {Ramani}}]{PhysRevD.99.035025}%
  \BibitemOpen
  \bibfield  {author} {\bibinfo {author} {\bibfnamefont {G.}~\bibnamefont
  {Benato}}, \bibinfo {author} {\bibfnamefont {A.}~\bibnamefont {Drobizhev}},
  \bibinfo {author} {\bibfnamefont {S.}~\bibnamefont {Rajendran}}, \ and\
  \bibinfo {author} {\bibfnamefont {H.}~\bibnamefont {Ramani}},\ }\href
  {\doibase 10.1103/PhysRevD.99.035025} {\bibfield  {journal} {\bibinfo
  {journal} {Phys. Rev. D}\ }\textbf {\bibinfo {volume} {99}},\ \bibinfo
  {pages} {035025} (\bibinfo {year} {2019})}\BibitemShut {NoStop}%
\bibitem [{\citenamefont {Dent}\ \emph {et~al.}(2022)\citenamefont {Dent},
  \citenamefont {Dutta}, \citenamefont {Jastram}, \citenamefont {Kim},
  \citenamefont {Kubik}, \citenamefont {Mahapatra}, \citenamefont {Rajendran},
  \citenamefont {Ramani}, \citenamefont {Thompson},\ and\ \citenamefont
  {Verma}}]{Dent:2021jnf}%
  \BibitemOpen
  \bibfield  {author} {\bibinfo {author} {\bibfnamefont {J.~B.}\ \bibnamefont
  {Dent}}, \bibinfo {author} {\bibfnamefont {B.}~\bibnamefont {Dutta}},
  \bibinfo {author} {\bibfnamefont {A.}~\bibnamefont {Jastram}}, \bibinfo
  {author} {\bibfnamefont {D.}~\bibnamefont {Kim}}, \bibinfo {author}
  {\bibfnamefont {A.}~\bibnamefont {Kubik}}, \bibinfo {author} {\bibfnamefont
  {R.}~\bibnamefont {Mahapatra}}, \bibinfo {author} {\bibfnamefont
  {S.}~\bibnamefont {Rajendran}}, \bibinfo {author} {\bibfnamefont
  {H.}~\bibnamefont {Ramani}}, \bibinfo {author} {\bibfnamefont
  {A.}~\bibnamefont {Thompson}}, \ and\ \bibinfo {author} {\bibfnamefont
  {S.}~\bibnamefont {Verma}},\ }\href {\doibase 10.1103/PhysRevD.105.015030}
  {\bibfield  {journal} {\bibinfo  {journal} {Phys. Rev. D}\ }\textbf {\bibinfo
  {volume} {105}},\ \bibinfo {pages} {015030} (\bibinfo {year} {2022})},\
  \Eprint {http://arxiv.org/abs/2105.07007} {arXiv:2105.07007 [hep-ph]}
  \BibitemShut {NoStop}%
\bibitem [{\citenamefont {Heeter}\ \emph {et~al.}(2021)\citenamefont {Heeter}
  \emph {et~al.}}]{lanlLaser}%
  \BibitemOpen
  \bibfield  {author} {\bibinfo {author} {\bibfnamefont {R.~F.}\ \bibnamefont
  {Heeter}} \emph {et~al.},\ }\href {\doibase 10.2172/1762842} {\  (\bibinfo
  {year} {2021}),\ 10.2172/1762842}\BibitemShut {NoStop}%
\bibitem [{\citenamefont {Yamaji}\ \emph {et~al.}(2018)\citenamefont {Yamaji},
  \citenamefont {Tamasaku}, \citenamefont {Namba}, \citenamefont {Yamazaki},\
  and\ \citenamefont {Seino}}]{Yamaji:2018ufo}%
  \BibitemOpen
  \bibfield  {author} {\bibinfo {author} {\bibfnamefont {T.}~\bibnamefont
  {Yamaji}}, \bibinfo {author} {\bibfnamefont {K.}~\bibnamefont {Tamasaku}},
  \bibinfo {author} {\bibfnamefont {T.}~\bibnamefont {Namba}}, \bibinfo
  {author} {\bibfnamefont {T.}~\bibnamefont {Yamazaki}}, \ and\ \bibinfo
  {author} {\bibfnamefont {Y.}~\bibnamefont {Seino}},\ }\href {\doibase
  10.1016/j.physletb.2018.05.068} {\bibfield  {journal} {\bibinfo  {journal}
  {Phys. Lett. B}\ }\textbf {\bibinfo {volume} {782}},\ \bibinfo {pages} {523}
  (\bibinfo {year} {2018})},\ \Eprint {http://arxiv.org/abs/1802.08388}
  {arXiv:1802.08388 [hep-ex]} \BibitemShut {NoStop}%
\bibitem [{\citenamefont {Battesti}\ \emph {et~al.}(2010)\citenamefont
  {Battesti}, \citenamefont {Fouch\'e}, \citenamefont {Detlefs}, \citenamefont
  {Roth}, \citenamefont {Berceau}, \citenamefont {Duc}, \citenamefont {Frings},
  \citenamefont {Rikken},\ and\ \citenamefont
  {Rizzo}}]{PhysRevLett.105.250405}%
  \BibitemOpen
  \bibfield  {author} {\bibinfo {author} {\bibfnamefont {R.}~\bibnamefont
  {Battesti}}, \bibinfo {author} {\bibfnamefont {M.}~\bibnamefont {Fouch\'e}},
  \bibinfo {author} {\bibfnamefont {C.}~\bibnamefont {Detlefs}}, \bibinfo
  {author} {\bibfnamefont {T.}~\bibnamefont {Roth}}, \bibinfo {author}
  {\bibfnamefont {P.}~\bibnamefont {Berceau}}, \bibinfo {author} {\bibfnamefont
  {F.}~\bibnamefont {Duc}}, \bibinfo {author} {\bibfnamefont {P.}~\bibnamefont
  {Frings}}, \bibinfo {author} {\bibfnamefont {G.~L. J.~A.}\ \bibnamefont
  {Rikken}}, \ and\ \bibinfo {author} {\bibfnamefont {C.}~\bibnamefont
  {Rizzo}},\ }\href {\doibase 10.1103/PhysRevLett.105.250405} {\bibfield
  {journal} {\bibinfo  {journal} {Phys. Rev. Lett.}\ }\textbf {\bibinfo
  {volume} {105}},\ \bibinfo {pages} {250405} (\bibinfo {year}
  {2010})}\BibitemShut {NoStop}%
\bibitem [{\citenamefont {Inada}\ \emph {et~al.}(2017)\citenamefont {Inada},
  \citenamefont {Yamazaki}, \citenamefont {Namba}, \citenamefont {Asai},
  \citenamefont {Kobayashi}, \citenamefont {Tamasaku}, \citenamefont {Tanaka},
  \citenamefont {Inubushi}, \citenamefont {Sawada}, \citenamefont {Yabashi},
  \citenamefont {Ishikawa}, \citenamefont {Matsuo}, \citenamefont {Kawaguchi},
  \citenamefont {Kindo},\ and\ \citenamefont
  {Nojiri}}]{PhysRevLett.118.071803}%
  \BibitemOpen
  \bibfield  {author} {\bibinfo {author} {\bibfnamefont {T.}~\bibnamefont
  {Inada}}, \bibinfo {author} {\bibfnamefont {T.}~\bibnamefont {Yamazaki}},
  \bibinfo {author} {\bibfnamefont {T.}~\bibnamefont {Namba}}, \bibinfo
  {author} {\bibfnamefont {S.}~\bibnamefont {Asai}}, \bibinfo {author}
  {\bibfnamefont {T.}~\bibnamefont {Kobayashi}}, \bibinfo {author}
  {\bibfnamefont {K.}~\bibnamefont {Tamasaku}}, \bibinfo {author}
  {\bibfnamefont {Y.}~\bibnamefont {Tanaka}}, \bibinfo {author} {\bibfnamefont
  {Y.}~\bibnamefont {Inubushi}}, \bibinfo {author} {\bibfnamefont
  {K.}~\bibnamefont {Sawada}}, \bibinfo {author} {\bibfnamefont
  {M.}~\bibnamefont {Yabashi}}, \bibinfo {author} {\bibfnamefont
  {T.}~\bibnamefont {Ishikawa}}, \bibinfo {author} {\bibfnamefont
  {A.}~\bibnamefont {Matsuo}}, \bibinfo {author} {\bibfnamefont
  {K.}~\bibnamefont {Kawaguchi}}, \bibinfo {author} {\bibfnamefont
  {K.}~\bibnamefont {Kindo}}, \ and\ \bibinfo {author} {\bibfnamefont
  {H.}~\bibnamefont {Nojiri}},\ }\href {\doibase
  10.1103/PhysRevLett.118.071803} {\bibfield  {journal} {\bibinfo  {journal}
  {Phys. Rev. Lett.}\ }\textbf {\bibinfo {volume} {118}},\ \bibinfo {pages}
  {071803} (\bibinfo {year} {2017})}\BibitemShut {NoStop}%
\bibitem [{\citenamefont {Beyer}\ \emph {et~al.}(2022)\citenamefont {Beyer},
  \citenamefont {Marocco}, \citenamefont {Bingham},\ and\ \citenamefont
  {Gregori}}]{PhysRevD.105.035031}%
  \BibitemOpen
  \bibfield  {author} {\bibinfo {author} {\bibfnamefont {K.~A.}\ \bibnamefont
  {Beyer}}, \bibinfo {author} {\bibfnamefont {G.}~\bibnamefont {Marocco}},
  \bibinfo {author} {\bibfnamefont {R.}~\bibnamefont {Bingham}}, \ and\
  \bibinfo {author} {\bibfnamefont {G.}~\bibnamefont {Gregori}},\ }\href
  {\doibase 10.1103/PhysRevD.105.035031} {\bibfield  {journal} {\bibinfo
  {journal} {Phys. Rev. D}\ }\textbf {\bibinfo {volume} {105}},\ \bibinfo
  {pages} {035031} (\bibinfo {year} {2022})}\BibitemShut {NoStop}%
\bibitem [{\citenamefont {B\"ahre}\ \emph {et~al.}(2013)\citenamefont {B\"ahre}
  \emph {et~al.}}]{Bahre:2013ywa}%
  \BibitemOpen
  \bibfield  {author} {\bibinfo {author} {\bibfnamefont {R.}~\bibnamefont
  {B\"ahre}} \emph {et~al.},\ }\href {\doibase 10.1088/1748-0221/8/09/T09001}
  {\bibfield  {journal} {\bibinfo  {journal} {JINST}\ }\textbf {\bibinfo
  {volume} {8}},\ \bibinfo {pages} {T09001} (\bibinfo {year} {2013})},\ \Eprint
  {http://arxiv.org/abs/1302.5647} {arXiv:1302.5647 [physics.ins-det]}
  \BibitemShut {NoStop}%
\bibitem [{\citenamefont {Anastassopoulos}\ \emph {et~al.}(2017)\citenamefont
  {Anastassopoulos} \emph {et~al.}}]{CAST:2017uph}%
  \BibitemOpen
  \bibfield  {author} {\bibinfo {author} {\bibfnamefont {V.}~\bibnamefont
  {Anastassopoulos}} \emph {et~al.} (\bibinfo {collaboration} {CAST}),\ }\href
  {\doibase 10.1038/nphys4109} {\bibfield  {journal} {\bibinfo  {journal}
  {Nature Phys.}\ }\textbf {\bibinfo {volume} {13}},\ \bibinfo {pages} {584}
  (\bibinfo {year} {2017})},\ \Eprint {http://arxiv.org/abs/1705.02290}
  {arXiv:1705.02290 [hep-ex]} \BibitemShut {NoStop}%
\bibitem [{\citenamefont {Gaillard}\ \emph {et~al.}(2018)\citenamefont
  {Gaillard}, \citenamefont {Gavela}, \citenamefont {Houtz}, \citenamefont
  {Quilez},\ and\ \citenamefont {Del~Rey}}]{Gaillard:2018xgk}%
  \BibitemOpen
  \bibfield  {author} {\bibinfo {author} {\bibfnamefont {M.~K.}\ \bibnamefont
  {Gaillard}}, \bibinfo {author} {\bibfnamefont {M.~B.}\ \bibnamefont
  {Gavela}}, \bibinfo {author} {\bibfnamefont {R.}~\bibnamefont {Houtz}},
  \bibinfo {author} {\bibfnamefont {P.}~\bibnamefont {Quilez}}, \ and\ \bibinfo
  {author} {\bibfnamefont {R.}~\bibnamefont {Del~Rey}},\ }\href {\doibase
  10.1140/epjc/s10052-018-6396-6} {\bibfield  {journal} {\bibinfo  {journal}
  {Eur. Phys. J. C}\ }\textbf {\bibinfo {volume} {78}},\ \bibinfo {pages} {972}
  (\bibinfo {year} {2018})},\ \Eprint {http://arxiv.org/abs/1805.06465}
  {arXiv:1805.06465 [hep-ph]} \BibitemShut {NoStop}%
\bibitem [{\citenamefont {Hook}\ \emph {et~al.}(2020)\citenamefont {Hook},
  \citenamefont {Kumar}, \citenamefont {Liu},\ and\ \citenamefont
  {Sundrum}}]{Hook:2019qoh}%
  \BibitemOpen
  \bibfield  {author} {\bibinfo {author} {\bibfnamefont {A.}~\bibnamefont
  {Hook}}, \bibinfo {author} {\bibfnamefont {S.}~\bibnamefont {Kumar}},
  \bibinfo {author} {\bibfnamefont {Z.}~\bibnamefont {Liu}}, \ and\ \bibinfo
  {author} {\bibfnamefont {R.}~\bibnamefont {Sundrum}},\ }\href {\doibase
  10.1103/PhysRevLett.124.221801} {\bibfield  {journal} {\bibinfo  {journal}
  {Phys. Rev. Lett.}\ }\textbf {\bibinfo {volume} {124}},\ \bibinfo {pages}
  {221801} (\bibinfo {year} {2020})},\ \Eprint
  {http://arxiv.org/abs/1911.12364} {arXiv:1911.12364 [hep-ph]} \BibitemShut
  {NoStop}%
\end{thebibliography}%

\onecolumngrid

\appendix

\section{Crystal structure}\label{app:crystals}
For convenience of the reader we repeat the standard discussion on the description of the lattice vector space for the crystals we have considered, much of which can be found in~\cite{warren} and other canonical literature. The $\vec{\alpha}_j$ describe the positions of each atom within the cell, while the basis vectors $\vec{a}_i$ describe the Bravais lattice. The linear combination of the two is used to translate anywhere on the lattice by stepping in integer multiples of these basis vectors;
\begin{equation}
    \vec{r}_i = n_1 \vec{a}_1 + n_2 \vec{a}_2 + n_3 \vec{a}_3 + \vec{\alpha}_i
\end{equation}
We can then introduce the reciprocal lattice, giving reciprocal lattice basis vectors $\vec{b}_i$ which satisfy $\vec{b}_i \cdot \vec{a}_j = 2\pi \delta_{ij}$. In general the transformations give
\begin{align}
\vec{b}_1 &= 2\pi \dfrac{\vec{a}_2 \times \vec{a}_3}{|\vec{a}_1 \cdot (\vec{a}_2 \times \vec{a}_3)|} \nonumber \\
\vec{b}_2 &= 2\pi \dfrac{\vec{a}_3 \times \vec{a}_1}{|\vec{a}_1 \cdot (\vec{a}_2 \times \vec{a}_3)|} \nonumber \\
\vec{b}_3 &= 2\pi \dfrac{\vec{a}_1 \times \vec{a}_2}{|\vec{a}_1 \cdot (\vec{a}_2 \times \vec{a}_3)|}
\end{align}
The reciprocal lattice basis vectors are used to construct the reciprocal lattice vector $\vec{G}$ that point along the surface normals of the scattering planes. In terms of integers $m_1$, $m_2$, and $m_3$, each scattering plane is defined;
\begin{equation}
    \vec{G} = m_1 \vec{b}_1 + m_2 \vec{b}_2 + m_3 \vec{b}_3
\end{equation}
Sometimes the integers $h,k,l$ are used instead, and in some contexts one can use this basis to express $\vec{G}$ as
\begin{equation}
    \vec{G}(hkl) = \dfrac{2\pi}{a} (h, k, l)
\end{equation}
The lattice constants, cell volumes, and basis vectors for a few examples (Ge, Si, CsI, and NaI) are listed in Table~\ref{tab:lattice}.

\begin{table*}[]
\begingroup
\renewcommand*{\arraystretch}{1.4}
    \centering
    \begin{tabularx}{0.9\textwidth}{|X|c|c|c|c|c|X}
    \hline
        Material & \thead{Lattice Constant\\$a$ (\AA)} & \thead{Cell Volume\\ $v_\textrm{cell}$ (\AA$^3$)} & Primitive Basis & Bravais Basis & $\vec{G}(m_1, m_2, m_3) \times a / 2\pi$\\
        \hline
         Ge (Diamond Cubic) & 5.657 & 181.0 & \multirow{2}{*}{\thead{$\vec{\alpha}_0 = (0,0,0)$\\$\vec{\alpha}_1 = \frac{a}{4}(1,1,1)$}} & \multirow{4}{*}{\thead{$\vec{a}_0 = \frac{a}{2}(0,0,0)$\\$\vec{a}_1 = \frac{a}{2}(1,0,1)$\\$\vec{a}_2 = \frac{a}{2}(0,1,1)$\\$\vec{a}_3 = \frac{a}{2}(1,1,0)$}} & \multirow{4}{*}{\thead{$(m_1 - m_2 + m_3,$\\$-m_1 + m_2 + m_3,$\\$m_1 + m_2 - m_3)$}}\\
         Si (Diamond Cubic) & 5.429 & 160.0 & & & \\
         \cline{1-4}
         CsI (FCC) & 4.503 & 91.3 & \multirow{2}{*}{\thead{$\vec{\alpha}_0 = (0,0,0)$\\$\vec{\alpha}_1 = \frac{a}{2}(1,1,1)$}} & & \\
         NaI (FCC) & 6.462 & 67.71 & & & \\
         \hline
    \end{tabularx}
\endgroup
    \caption{Lattice information for typical crystal detector technologies.}
    \label{tab:lattice}
\end{table*}

\section{Derivation of the Event Rate}
\label{app:derivation}
Let $f(\vec{k},\vec{k}^\prime)$ be the Primakoff scattering matrix element for a single atomic target, for an incoming ALP 3-momentum $\vec{k}$ and outgoing $\gamma$ 3-momentum $\vec{k}^\prime$;
\begin{equation}
    f = \mathcal{M}_\textrm{free} F_A (q)
\end{equation}
where $\mathcal{M}_\textrm{free}$ is the single-atomic scattering amplitude with the angle of scattering defined by $\vec{k}_a \cdot \vec{k}_\gamma = E_\gamma k \cos2\theta$, averaged over spins and taken in the limit $k \gg m_a$, $m_N \gg k,E_\gamma$;
\begin{equation}
    \braket{\mathcal{M}_\textrm{free}} = \dfrac{8 e^2 g_{a\gamma}^2}{q^4} E_\gamma^2 m_N^2 k^2 \sin^2 2\theta 
\end{equation}
We sum over the $N$ scattering centers in a crystal;
\begin{equation}
    \mathcal{M}(\vec{k},\vec{k}^\prime) = \sum_{j=1}^N f_j(\vec{k},\vec{k}^\prime) e^{i(\vec{k}^\prime - \vec{k})\cdot \vec{r}_j}
\end{equation}
where $e^{i(\vec{k}^\prime - \vec{k})\cdot \vec{r}_j}$ is a phase factor that comes from assuming plane wave solutions for the in and out states. The position vector $\vec{r}_j$ can be expressed in terms of the Bravais lattice basis vectors and the primitive basis vectors for each unit cell of the crystal. For germanium crystal with lattice constant $a$, we have \textit{primitive basis vectors}
\begin{align}
    \vec{\alpha}_0 &= (0,0,0) \nonumber \\
    \vec{\alpha}_1 &= \frac{a}{4} (1,1,1)
\end{align}
while the \textit{basis vectors of the Bravais lattice} are described by $\vec{a}_1$, $\vec{a}_2$, and $\vec{a}_3$;
\begin{align}
    \vec{a}_1 &= \frac{a}{2}(0,1,1) \nonumber \\
    \vec{a}_2 &= \frac{a}{2} (1,0,1) \nonumber \\
    \vec{a}_3 &= \frac{a}{2} (1,1,0)
\end{align}
we can represent any scattering site as a linear combination of the $a$'s and either the first or second primitive;
\begin{align}
    \vec{r}_{i,0} &= \vec{R}_i  + \vec{\alpha}_0 = n_1 \vec{a}_1 + n_2 \vec{a}_2 + n_3 \vec{a}_3 + \vec{\alpha}_0 \\
    \vec{r}_{i,1} &= \vec{R}_i  + \vec{\alpha}_1 = n_1 \vec{a}_1 + n_2 \vec{a}_2 + n_3 \vec{a}_3 + \vec{\alpha}_1
\end{align}
where the index $i$ maps to a unique combination ($n_1, n_2, n_3$).
If we square this, we get
\begin{equation}
     \mid\mathcal{M}(\vec{k},\vec{k}^\prime)\mid^2 = \sum_{i=1}^N \mid f_i\mid^2 + \sum_{j\neq i}^N \sum_{i=1}^N f_j^\dagger f_i e^{-i\vec{q}\cdot(\vec{r}_i - \vec{r}_j)}
\end{equation}
taking $\vec{q} \equiv \vec{k} - \vec{k}^\prime$. Rewriting in terms of a sum over $N_c$ cells and the cell primitives, the coherent part (second term) is
\begin{equation}
     \mid\mathcal{M}(\vec{k},\vec{k}^\prime)\mid^2 = \sum_{j\neq i}^{N_c} \sum_{i=1}^{N_c} \sum_{\mu = 0}^1\sum_{\nu = 0}^1 f_j^\dagger f_i e^{-i\vec{q}\cdot(\vec{R}_i - \vec{R}_j + \vec{\alpha}_\mu - \vec{\alpha}_\nu)}
\end{equation}
When the Laue condition is met, we have $\vec{q} = \vec{G}$ and $\vec{G}\cdot \vec{R}_i$ is a $2\pi$ integer multiple;
\begin{equation}
 |\mathcal{M}|^2 \equiv \sum_{j\neq i}^{N_c} \sum_{i=1}^{N_c} \sum_{\mu,\nu = 0}^1 f_j^\dagger f_i e^{-i\vec{G}\cdot(\vec{\alpha}_\mu - \vec{\alpha}_\nu)}
    \label{eq:cohsum}
\end{equation}
Now we can factorize the sum over primitives, and since we are considering a monoatomic crystal we can also take the $f_i = f_j$, simplifying things;
\begin{equation}
 |\mathcal{M}|^2 = N_c^2 f^\dagger f \sum_{\mu,\nu = 0}^1  e^{-i\vec{G}\cdot(\vec{\alpha}_\mu - \vec{\alpha}_\nu)} 
    \label{eq:cohsum2}
\end{equation}
In Eq.~\ref{eq:cohsum2} the structure function can be substituted, which is nothing but the sum over primtives;
\begin{equation}
    S(\vec{G}) = \sum_\mu e^{i \vec{G}\cdot\alpha_\mu}
\end{equation}
and we have no need for a species index $j$ on $S_j(\vec{G})$ since we only have one atomic species, but it is trivial to extend this derivation to include it - we just need to add another index to the primitive basis vectors and sum over it. With this identification and also taking $f^\dagger f = |\mathcal{M}_\textrm{free}|^2 F^2_A (\vec{G})$, we have

\begin{align}
 |\mathcal{M}|^2 &= N_c^2 |\mathcal{M}_\textrm{free}|^2 |F_A (\vec{G}) S(\vec{G})|^2
\end{align}

Now let's write down the cross section.
\begin{equation}
    d\sigma = \dfrac{1}{4 E_a m_N v_a} |\mathcal{M}|^2 \dfrac{d^3 k^\prime}{(2\pi)^3 2E_\gamma}\dfrac{d^3 p^\prime}{(2\pi)^3 2E_{p^\prime}} (2\pi)^4 \delta^4 (k + p - k^\prime - p^\prime)
\end{equation}
Taking the ALP velocity $v_a = 1$, momentum transfer minimal such that $E_{p^\prime} = m_N$, and integrating out the $\delta^3$ we get
\begin{equation}
    d\sigma = \dfrac{1}{64 \pi^2 E_a E_\gamma m_N^2} |\mathcal{M}|^2  d^3 k^\prime \delta(E_a - E_\gamma)
\end{equation}
Performing a change of variables to $d^3k^\prime \to d^3q$ (since $q = k - k^\prime$ and $k$ is fixed), we would integrate this over $\vec{q}$. Since we have $\vec{q} = \vec{G}$ at this stage, we should replace the integral with a sum;
\begin{equation}
    \int d^3 q \to \frac{(2\pi)^3}{V} \sum_{\vec{G}}
\end{equation}
The event rate formula is constructed from a convolution of the detector response, axion flux $\Phi_a$, and cross section;
\begin{equation}
    \dfrac{dN}{dt} = \int_{E_1}^{E_2} dE_{ee} \int_0^\infty dE_a  \frac{(2\pi)^3}{V}\sum_{\vec{G}} \dfrac{d\Phi_a}{d E_a}   \dfrac{1}{64 \pi^2 E_a E_\gamma m_N^2} |\mathcal{M}|^2 \delta(E_a - E_\gamma) \cdot \bigg( \dfrac{1}{\Delta \sqrt{2\pi}} e^{-(E_{ee} - E_\gamma)^2/2\Delta^2} \bigg)
\end{equation}
Putting in the definition of $|\mathcal{M}|^2$ that we worked out and substituting the free Primakoff cross section, integrating over the energy delta function (and identifying $E_a = E_\gamma = E$ for simplicity), and integrating over $dE_{ee}$ we get
\begin{equation}
\label{eq:event_rate_app}
    \dfrac{dN}{dt} = \dfrac{(2\pi)^3 e^2 g_{a\gamma}^2}{8 \pi^2} \dfrac{V}{v_\text{cell}^2} \sum_{\vec{G}} \dfrac{d\Phi_a}{dE} \dfrac{k^2 \sin^2 (2\theta)}{|\vec{G}|^4} |F_A(\vec{G})S(\vec{G})|^2 \mathcal{W}(E_1, E_2, E)
\end{equation}
This is almost identical to the rate in ref.~\cite{Cebrian:1998mu}, which uses a different definition of the atomic form factor up to a factor of $\frac{q^2}{e k^2}$. After some algebra, the event rate in Eq.~\ref{eq:event_rate_app} is still different than that given in ref.~\cite{Cebrian:1998mu} up to a factor of $4\sin^2(\theta)$. However, the event rate formula derived here is consistent with the calculation performed in refs.~\cite{Li:2015tsa, Li:2015tyq}. After rederiving the coherent sum using the replacements in Eqns.~\ref{eq:dielectric}-\ref{eq:abs_sum}, the event rate becomes
\begin{equation}
\label{eq:event_rate_app_abs}
    \dfrac{dN}{dt} = \dfrac{(2\pi)^3 e^2 g_{a\gamma}^2}{8 \pi^2} \dfrac{V}{v_\text{cell}^2} \sum_{\vec{G}} I(\vec{k},\vec{G}) \dfrac{d\Phi_a}{dE} \dfrac{k^2 \sin^2 (2\theta)}{|\vec{G}|^4} |F_A(\vec{G})S(\vec{G})|^2 \mathcal{W}(E_1, E_2, E)
\end{equation}

\section{Solar Axion Flux}
\begin{figure}
    \centering
    \includegraphics[width=0.8\textwidth]{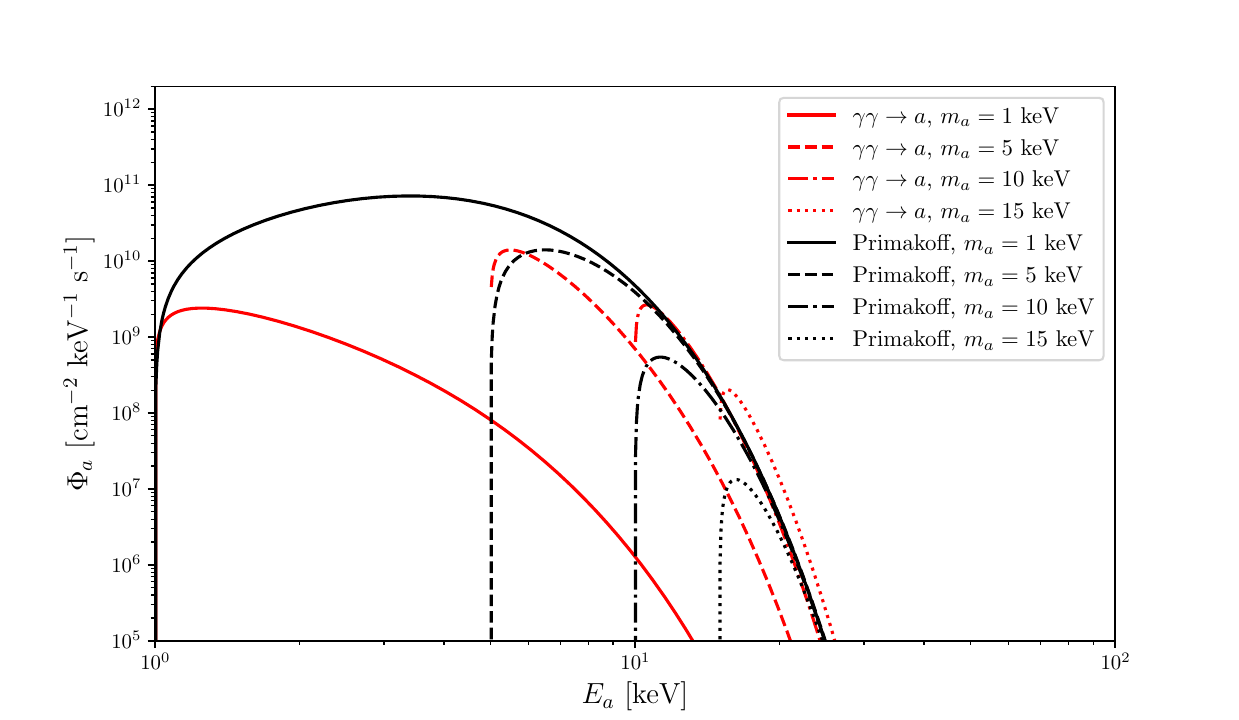}
    \caption{Solar Axion fluxes produced from Primakoff scattering ($\gamma Z \to a Z$) and coalescence ($\gamma \gamma \to a$) in the sun.}
    \label{fig:solar_fluxes}
\end{figure}

We use the parameterization appearing in ref.~\cite{DiLella:2000dn} for massive axion production in the sun; the flux parameterizations are repeated here for convenience
\begin{align}
\label{eq:solar_fluxes}
    \dfrac{d\Phi_{\gamma\to a}}{dE_a} &= \dfrac{4.20\cdot 10^{10}}{\textrm{cm}^{-2}\textrm{s}^{-1}\textrm{keV}^{-1}} \bigg(\dfrac{g_{a\gamma}}{10^{-10}\textrm{GeV}^{-1}}\bigg)^2 \dfrac{E_a p_a^2}{e^{E_a/1.1} -0.7} (1 + 0.02 m_a)\\
    \dfrac{d\Phi_{\gamma\gamma\to a}}{dE_a} &= \dfrac{1.68\cdot 10^9}{\textrm{cm}^{-2}\textrm{s}^{-1}\textrm{keV}^{-1}} \bigg(\dfrac{g_{a\gamma}}{10^{-10}\textrm{GeV}^{-1}}\bigg)^2 m_a^4 p_a \bigg(1 + 0.0006 E_a^3 + \frac{10}{E_a^2 + 0.2}\bigg) e^{-E_a}
\end{align}
where $\Phi_{\gamma \to a}$ is the Primakoff solar flux and $\Phi_{\gamma\gamma\to a}$ is the flux resulting from resonant photon coalescence, both in units of $\textrm{cm}^{-2}\textrm{s}^{-1}\textrm{keV}^{-1}$, given for axion energy and momentum $E_a$ and $p_a$ in keV, and for the coupling $g_{a\gamma}$ in GeV$^{-1}$. The solar axion flux from photon coalescence and Primakoff conversion is shown in Fig.~\ref{fig:solar_fluxes} for several benchmark axion masses.

\section{Utilizing \texttt{DarkARC}/\texttt{DarkART} for Calculation of the Absorptive Form Factor}\label{app:darkarc}
\begin{figure}[hb!]
    \centering
    \includegraphics[width=0.6\textwidth]{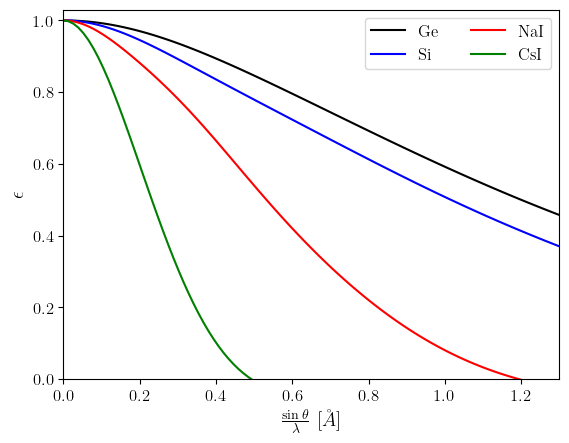}
    \caption{Borrmann parameter $\epsilon$ as a function of the momentum transfer $\sin\theta / \lambda = |\vec{G}|/4\pi$ for several crystal materials.}
    \label{fig:borrmann_by_mat}
\end{figure}
Wagenfield's form factor for the anomalous dispersion of X-rays with incoming and outgoing momenta and polarizations $\vb{k}$, $\polvec_0$, $\vb{k}^\prime$, $\polvec_0^\prime$  is~\cite{wagenfield1986}
\begin{align}
    \Delta f^{\prime\prime} &= \dfrac{\pi \hbar^2}{m_e} \bigg( \int \psi_f^*(\vb{r}) \polvec_0 \cdot \vb{\nabla} e^{i \vb{k}\cdot\vb{r}} \psi_i(\vb{r}) d^3 r \bigg) \bigg( \int \psi_f(\vb{r}) \polvec^\prime_0 \cdot \vb{\nabla} e^{-i \vb{k}^\prime\cdot\vb{r}} \psi_i^*(\vb{r}) d^3 r \bigg)
\end{align}
Applying the gradient and expanding, we get some terms proportional to $\polvec_0 \cdot \vb{k}$ which vanish, leaving us with
\begin{align}
    \Delta f^{\prime\prime} &= \dfrac{\pi \hbar^2}{m_e} \bigg( \polvec_0 \cdot \int \psi_f^*(\vb{r}) e^{i \vb{k}\cdot\vb{r}} \vb{\nabla}\psi_i(\vb{r}) d^3 r \bigg)  \bigg(  \polvec^\prime_0 \cdot \int \psi_f(\vb{r}) e^{-i \vb{k}^\prime\cdot\vb{r}} \vb{\nabla}\psi_i^*(\vb{r}) d^3 r \bigg)
\end{align}
Referring to Catena \textit{et al}~\cite{Catena:2019gfa}, we can then apply the definition of the vectorial form factor (eq B18, but with some changes made to keep the notation more consistent),
\begin{equation}
    \vb{f}_{1\to2}(\vb{q}) = \int d^3 r \psi^*_f (\vb{r}) e^{i \vb{q}\cdot\vb{r}} \frac{i \vb{\nabla}}{m_e} \psi_i (\vb{r}).
\end{equation}
Here the final state and initial state wave functions have quantum numbers $i = n,\ell,m$ and $f = p^\prime,\ell^\prime, m^\prime$ where $p^\prime$ is the final state electron momentum, and $\{n,\ell,m\},\{\ell^\prime,m^\prime\}$ are the initial and final quantum numbers, respectively. Applying this definition, we have
\begin{align}
    \Delta f^{\prime\prime} &= \dfrac{\pi \hbar^2}{m_e} \bigg( \polvec_0 \cdot (-i m_e) \vb{f}_{1\to2}(\vb{k})  \bigg) \bigg(  \polvec^\prime_0 \cdot (i m_e) \vb{f}^*_{1\to2}(\vb{k}^\prime) \bigg) \nonumber \\
    &=\pi \hbar^2 m_e \bigg(\polvec_0 \cdot \vb{f}_{1\to2}(\vb{k})\bigg) \bigg(\polvec_0^\prime \cdot \vb{f}^*_{1\to2}(\vb{k}^\prime)\bigg)
\end{align}
If our photons are unpolarized, then we can take a sum over the helicity states, giving the completeness relation $\sum_s (\polvec_0(s))_i (\polvec_0^\prime(s))_j = \delta_{ij}$. Taking $\vb{k}^\prime = \vb{k} - \vb{q}$, this reduces the polarization-summed imaginary form factor to
\begin{equation}
    \Delta f^{\prime\prime}(\vb{k},\vb{q}) = \pi \hbar^2 m_e \bigg(\vb{f}_{1\to2}(\vb{k})  \cdot \vb{f}^*_{1\to2}(\vb{k} - \vb{q})\bigg)
\end{equation}

\end{document}